\documentclass[floatfix, twocolumn, 10pt, aps, prd, showpacs, amsmath, amssymb, superscriptaddress]{revtex4-1}
\usepackage{graphicx}
\usepackage{hyperref}
\usepackage{subeqnarray}
\usepackage{color}
\usepackage[normalem]{ulem}
\usepackage{url}
\usepackage{multirow}
\hypersetup{colorlinks=false}
\hypersetup{linktocpage}
\definecolor{darkblue}{RGB}{16,78,139}
\definecolor{darkgreen}{rgb}{0.04, 0.7, 0.2}
\newcommand{\beq}{\begin{equation}}
\newcommand{\eeq}{\end{equation}}
\newcommand{\beqn}{\begin{eqnarray}}
\newcommand{\eeqn}{\end{eqnarray}}
\newcommand{\beqs}{\begin{subeqnarray}}
\newcommand{\eeqs}{\end{subeqnarray}}
\newcommand{\nn}{\nonumber}

\begin{document}
\title{Quasinormal modes of the polytropic hydrodynamic vortex}

\author{Leandro A. Oliveira}
\email{leandro.oliveira@york.ac.uk}
\affiliation{Department of Mathematics, University of York, YO10 5DD Heslington, York, United Kingdom.}

\author{Vitor Cardoso}
\email{vitor.cardoso@ist.utl.pt}
\affiliation{CENTRA, Departamento de F\'\i sica, Instituto Superior T\'ecnico, Universidade de Lisboa--UL, Av. Rovisco Pais 1, 1049 Lisboa, Portugal.}
\affiliation{Faculdade de F\'{\i}sica, Universidade Federal do Par\'a, 66075-110, Bel\'em, Par\'a, Brazil.}
\affiliation{Perimeter Institute for Theoretical Physics Waterloo, Ontario N2J 2W9, Canada.}

\author{Lu\'is C. B. Crispino}
\email{crispino@ufpa.br}
\affiliation{Faculdade de F\'{\i}sica, Universidade 
Federal do Par\'a, 66075-110, Bel\'em, Par\'a, Brazil.}

\date{\today}

\begin{abstract}
Analogue systems are a powerful instrument to investigate and understand in a controlled setting many general-relativistic effects.
Here, we focus on superradiant-triggered instabilities and quasi-normal modes. We consider a compressible hydrodynamic vortex characterized by a polytropic equation of state, the {\it polytropic hydrodynamic vortex}, a purely circulating system with an ergoregion but no event horizon. We compute the quasinormal modes of this system numerically with different methods, finding excellent agreement between them. When the fluid velocity is larger than the speed of sound, an ergoregion appears in the effective spacetime, triggering an ``ergoregion instability.'' We study the details of the instability for the polytropic vortex, and in particular find analytic expressions for the marginally-stable configuration.
\end{abstract}


\maketitle

\section{Introduction}
\label{sec-Introduction} 
Black holes are an important component of our Universe, thought to play a role in the dynamics of galaxies and in star formation.
They have also come to play a prominent role in high-energy physics, and even fundamental physics as they are simultaneously an ``elementary particle'' of gravitation and an entity where classical and quantum effects are intertwined through Hawking radiation.
Unfortunately, many of these features and properties have the undesirable consequence that black holes are ``hard to see''. It is therefore very convenient to use analogue setups where these properties are present, but which can be manipulated in the laboratory. One such example are acoustic holes, idealized models in fluid dynamics mimicking the behavior of a curved spacetime~\cite{Unruh:1980cg,Visser:1997ux, AMproc}. 

Acoustic holes have, for instance, been used to understand wave scattering in more general curved spacetimes~\cite{Crispino:2007zz, Oliveira:2010zzb,Dolan:2009zza,Dolan:2011zza,Dolan:2012yc} as well as the impact of UV cutoffs in the Hawking radiation in reasonable completions of General Relativity~\cite{Barcelo:2005fc}. One specially interesting feature of curved spacetimes in General Relativity is the ergoregion, bounded by an infinite-redshift surface which represents the static limit where it's impossible to remain at rest with respect to distant inertial observers. Negative-energy states are possible within the ergoregion, leading to superradiant effects
if horizons are present and to instabilities otherwise~\cite{Brito:2015oca}.

The purpose of this work is to explore the phenomenology of superradiant instabilities in analogue models.
We focus on the hydrodynamic vortex~\cite{Fischer:2001jz,Visser:2004zs,Slatyer:2005ty}, a purely circulating compressible system, and we will
work with a polytropic equation of state~\cite{no-horizon}. 

Here we show that the polytropic hydrodynamic vortex is unstable under linearized perturbations, and that the appearance of these instabilities is directly related to the presence of an ergoregion and absence of an event horizon, the so-called ergoregion instability~\cite{Friedman:1978wla,Brito:2015oca}. This result is an generalization of a previous result (obtained for an incompressible system in \cite{Oliveira:2014oja}) for compressible systems that satisfy a polytropic equation of state and that describe a wide-class of thermodynamical processes, as, e.g., isentropic processes~\cite{Cherubini:2011zza, Cherubini:2013iea, Horedt, Turns}. Here we focus on values of polytropic index $N_{\rm p}$ describing isentropic processes (also adiabatic processes), with a compatible experimental setup in perfect gases.

The remainder of this paper is structured as follows. In Sec.~\ref{sec-Metric} we describe the spacetime of the polytropic hydrodynamic vortex. In Sec.~\ref{sec-Perturbations} we study the perturbations of the polytropic hydrodynamic vortex using descriptions in the time and frequency domains. In Sec.~\ref{sec-Numerical} we obtain the quasinormal mode (QNM) frequencies of the polytropic hydrodynamic vortex using the method of lines (MOL), direct integration (DI) and the continued fraction (CF) method. In Sec.~\ref{sec-Results} we validate and comment our results comparing the QNM frequencies obtained via MOL, DI and CF methods. Furthermore, we investigate the static (marginally-stable) resonances of the polytropic hydrodynamic vortex, studying this system in the regime between stability and instability. We conclude with a brief discussion in Sec.~\ref{sec-Conclusion}.

\section{The polytropic hydrodynamic vortex}
\label{sec-Metric}
The (effective) spacetime of the polytropic hydrodynamic vortex is produced by an irrotational, barotropic and purely circulating fluid characterized by a polytropic equation of state. The line element of this system may be written as
\beq
ds^2=\frac{\rho}{c_{\rm s}}\left[-c_{\rm s}^2dt^2+\left(rd\theta -v_\theta dt \right)^2+dr^2+dz^2\right],
\label{vortex}
\eeq
where $\rho$ is the mass density, $v_\theta$ is the angular component of the flow velocity $\vec{v}$, i.e., $\vec{v}=v_\theta \hat{\theta}$, and $c_{\rm s}$ is the speed of sound, which may be defined as
\beq
c_{\rm s} \equiv \sqrt{\frac{dP}{d\rho}},
\label{speed}
\eeq
assuming that the fluid is barotropic, i.e., 
\beq
P=P(\rho), 
\label{state_1}
\eeq
where $P$ is the hydrostatic pressure. Note that the quantities $\rho$, $P$, $\vec{v}$ and $c_{\rm s}$ are given by the local properties of the unperturbed fluid flow~\cite{Visser:1997ux}.

We may obtain expressions for $\rho$, $c_{\rm s}$ and $v_\theta$ assuming that the fluid flow is irrotational 
\beq
\nabla \times \vec{v}=0,
\label{irrot}
\eeq
and that it satisfies the Euler equation, i.e,
\beq
\frac{\partial \vec{v}}{\partial t} +\frac{1}{2}\nabla v^{2} +\frac{\nabla P}{\rho}=0.
\label{conser}
\eeq
Considering that density $\rho$, pressure $P$ and angular flow velocity $v_\theta $ are functions of the radial coordinate only, i.e., $\rho = \rho(r)$, $P = P(r)$, $v_\theta = v_\theta(r)$, we obtain, respectively, from Eqs.~(\ref{irrot}) and~(\ref{conser}), the following expressions
\beqs \label{conser1}
\slabel{curl1} r\frac{dv_\theta}{dr}+v_\theta=0,\\
\slabel{euler1} \frac{v_\theta^{2}}{r}-\frac{1}{\rho} \frac{dP}{dr}=0.
\eeqs

From Eqs.~(\ref{conser1}), it follows that
\beqn 
 &&v_\theta = \frac{C}{r},\label{velo}\\
 &&\frac{dP}{dr}-\frac{C^{2}}{r^{3}}\rho=0,
\label{state}
\eeqn
where $C$ is a constant related to the circulation of the fluid~\cite{Fischer:2001jz}.

We may solve Eq.~(\ref{state}) using an expression that denotes the relation between $P$ and $\rho$, i.e., an equation of state [cf. Eq.~(\ref{state_1})]. Here we use a polytropic equation of state, namely
\beq
P=k_{\rm p}\rho^{1+1/N_{\rm p}},
\label{poly}
\eeq
where $k_{\rm p}$ is the polytropic constant and $N_{\rm p}$ is a constant called polytropic index~\cite{Horedt}. 

Certain thermodynamical processes, in which one of
quantities of the system remains constant, can be described by specific values of the polytropic index $N_{\rm p}$~\cite{Horedt, Turns}, namely: (i) isobaric (constant pressure) processes, described by polytropic index $N_{\rm p}=-1$; (ii) isometric (constant volume) and isopycnic (constant density) processes, described by polytropic index $N_{\rm p}=0$; (iii) isothermal (constant temperature) processes, described by polytropic index $N_{\rm p}=\pm \infty$; and
(iv) isentropic (constant entropy) processes, described by polytropic index $N_{\rm p}=1/(\lambda-1)$. The quantity $\lambda$ is the so-called {\it specific heat ratio}, being given by $\lambda=c_p/c_v$, where $ c_p$ and $c_v$ are, respectively, the specific heat at constant pressure and the specific heat at constant volume, for a perfect gas~\cite{Horedt}. 

Substituting Eq.~(\ref{poly}) into Eq.~(\ref{state}), we obtain the following first-order differential equation:
\beq
k_{\rm p}\left(1+\frac{1}{N_{\rm p}}\right)\rho^{1/N_{\rm p}} \frac{d\rho}{dr}-\frac{C^{2}}{r^{3}}\rho=0,
\label{eq_density}
\eeq
whose solution may be written as
\beq
\rho(r) =\rho_\infty\left( 1-\frac{r_{\rm c}^{2}}{r^{2}}\right)^{N_{\rm p}},
\label{density}
\eeq
where $\rho_\infty$ is the density at $r\rightarrow \infty$. Note that the density goes to zero and becomes non-physical at a critical radius~\cite{non-physical} defined by
\beq
r_{\rm c}\equiv \frac{|C|}{\sqrt{2K_{\rm p}(N_{\rm p}+1)}}.
\label{critical}
\eeq
Here we defined $K_{\rm p} \equiv k_{\rm p}\rho_\infty^{1/N_{\rm p}}$. Therefore, the polytropic hydrodynamic vortex has an essential singularity at the critical radius $r=r_{\rm c}$, denoting that this spacetime, as well as Kerr spacetime, has a singularity with non-pointlike structure \cite{Cherubini:2011zza}.

Using the equation of state given by Eq.~(\ref{poly}) and the definition of the speed of sound given by Eq.~(\ref{speed}), we may obtain an expression to the speed of sound, as a function of the density, for a polytropic fluid, i.e.,
\beq
c_{\rm s} = \sqrt{k_{\rm p} \left( 1+\frac{1}{N_{\rm p}}\right)\rho^{1/N_{\rm p}}}.
\label{speed2}
\eeq
Using Eqs.~(\ref{density}) and~(\ref{speed2}), we obtain get
\beq
c_{\rm s}(r)=c_{\rm s\infty}\sqrt{ 1-\frac{r_{\rm c}^{2}}{r^{2}}},
\label{speed3}
\eeq
where $c_{\rm s\infty}$ is the speed of sound at $r\rightarrow \infty$, given by
\beq
c_{\rm s\infty} = \sqrt{K_{\rm p} \left( 1+\frac{1}{N_{\rm p}}\right)}.
\label{csinf}
\eeq
As it happens to the density [cf. Eq.~\eqref{density}], the speed of sound [cf. Eq.~\eqref{speed3}] becomes zero and non-physical at the critical radius $r_{\rm c}$~\cite{non-physical} [see Eq.~(\ref{critical})]. 

The local Mach number is defined as being the ratio between absolute value of the flow velocity and speed of sound, i.e.,  $M\equiv|\vec{v}|/c_{\rm s}$. Using the local Mach number as a parameter, we may define a supersonic (subsonic) flow as $M > 1$ ($M < 1$). For $M=1$ (or, explicitly, $|\vec{v}|=c_{\rm s}$), we may write the outer boundary of the ergoregion, $r_{\rm e}$, as 
\beq
 r_{\rm e} \equiv \frac{|C|\sqrt{2N_{\rm p}+1}}{\sqrt{2K_{\rm p} \left( N_{\rm p}+1\right)}}.
 \label{re}
\eeq 
As seen from line element given by Eq.~\eqref{vortex}, the polytropic hydrodynamic vortex has no coordinate singularities, i.e., it has no event horizon, but it has an ergoregion with outer boundary at $r=r_{\rm e}$.

Dividing Eq.~(\ref{re}) by Eq.~(\ref{critical}) we find the following ratio between the radius of the outer boundary of the ergoregion $r_{\rm e}$ and critical radius $r_{\rm c}$:
\beq
\frac{r_{\rm e}}{r_{\rm c}} = \sqrt{2N_{\rm p}+1}.
\label{re_rc}
\eeq

From Eq.~\eqref{re_rc}, for $N_{\rm p} \geq 0$, we find that $r_{\rm e} \geq r_{\rm c}$, i.e., the outer boundary of the ergoregion $r_{\rm e}$ is located outside the outer boundary of the critical radius $r_{\rm c}$. For $-1/2 \leq N_{\rm p} < 0$, the outer boundary of the ergoregion $r_{\rm e}$ is located inside of the outer boundary of the critical radius $r_{\rm c}$, i.e.,  $r_{\rm e} < r_{\rm c}$. For $N_{\rm p} < -1/2$, the outer boundary of the ergoregion $r_{\rm e}$ is complex, and therefore non-physical.

Basically, we are interested in values of the polytropic index such that $r_{\rm e}$ is real and $r_{\rm e} > r_{\rm c}$, i.e., $N_{\rm p} > 0$. As seen in Eq.~\eqref{critical}, the case in which $N_{\rm p}=0$ gives us $r_{\rm c} \rightarrow 0$, therefore we will not consider this case here. The $N_{\rm p}=0$ case [which describe an isopycnic (constant density) processes] was analyzed in Ref.~\cite{Oliveira:2014oja}. Here we focus our attention in the case of the polytropic index describing isentropic processes.

Essentially, an isentropic process is always an adiabatic process too, i.e., in addition to the entropy of the system remain constant, the process occurs without transfer of heat energy throughout~\cite{Horedt}. The main motivation to study isentropic (adiabatic) processes here is based on fact that the propagation of sound waves in a gas is fundamentally an adiabatic process~\cite{Turrell, Laplace}.

We now assume the polytropic hydrodynamic vortex to be composed by a perfect gas, described by the equation~\cite{Horedt}
\beq
P=\frac{RT}{\mu}\rho,
\label{perfect}
\eeq
where $R=8.314462\, \text{J}\, \text{K}^{-1}\, \text{mol}^{-1}$ is the perfect gas constant, $T$ is the temperature of the gas, and $\mu$ is the mean molecular weight of the gas. In this case, from Eqs.~\eqref{poly},~\eqref{density} and~\eqref{perfect}, we may write expressions for the pressure and temperature of the gas, as functions of $r$, given, respectively, by
\beq
P(r)=P_{\infty}\left( 1-\frac{r_{\rm c}^{2}}{r^{2}}\right)^{N_{\rm p}+1},
\label{pressure}
\eeq
where $P_{\infty}$ is the pressure at $r\rightarrow \infty$, and
\beq
T(r)=T_{\infty}\left( 1-\frac{r_{\rm c}^{2}}{r^{2}}\right),
\label{temperature}
\eeq
where $T_{\infty}$ is the temperature at $r\rightarrow \infty$. Pressure and temperature, as the density and speed of sound, becomes zero and non-physical at critical radius $r_{\rm c}$~\cite{non-physical} (see Eq.~(\ref{critical})).

In Table~\ref{tab-values} we exhibit estimates of the polytropic index $N_{\rm p}$, of the constant $K_{\rm p}$ [which, using Eqs.~\eqref{perfect} and~\eqref{temperature}, may be written as $K_{\rm p}=RT_\infty/\mu$], of the critical radius $r_{\rm c}$, of the density at infinity $\rho_\infty$, of the speed of sound at infinity $c_{\rm s \infty}$, and of the ratio $c_{\rm s \infty}/r_{\rm c}$, for some gases.
\begin{table*}[htpb!]
\caption{Estimates of the polytropic index $N_{\rm p}$, of the constant $K_{\rm p}$, of the critical radius $r_{\rm c}$, of the density at infinity $\rho_\infty$, speed of sound at infinity $c_{\rm s \infty}$, and of the ratio $c_{\rm s \infty}/r_{\rm c}$, for some gases, obtained considering the temperature $T_\infty =  288.15 \, \text{K} $ and the pressure $P_\infty = 101325\, \text{Pa}$ (estimates of $N_{\rm p}$, $K_{\rm p}$ and $\mu$ were obtained from data extracted from Ref.~\cite{Perry}). Here we choose the circulation $C=0.5\,\,\text{m}^2/\text{s}$. In parentheses are indicated the units and the order of magnitude of each quantity.} 
\begin{center}
\begin{tabular}{c c c c c c c}
\hline\hline
 \multicolumn{1}{c}{Gas} & \multicolumn{1}{c}{$N_{\rm p}$} &  \multicolumn{1}{c}{$K_{\rm p} \, \left(10^4 \,\text{m}^2/\text{s}^2\right)$}   & \multicolumn{1}{c}{$r_{\rm c} \left(10^{-4} \,\text{m}\right)$} & \multicolumn{1}{c}{$\rho_\infty (\text{kg/m$^3$})$}  & \multicolumn{1}{c}{$ c_{\rm s \infty} \left(\text{m}/\text{s}\right)$}  & \multicolumn{1}{c}{$c_{\rm s \infty}/r_{\rm c} \left(10^{5} \,\text{s}^{-1}\right)$}\\
\hline
Argon ($\text{Ar}$)            \quad  & $1.49$  &  $ 5.99853  $  & $9.14814  $  &  $ 1.68916  $  & $ 316.613 $ & $ 3.46095 $ \\
Nitrogen ($\text{N}_2$)        \quad  & $2.47$  &  $ 8.55036  $  & $6.49080  $  &  $ 1.18504  $  & $ 346.584 $ & $ 5.33962 $ \\
Carbon dioxide ($\text{CO}_2$) \quad \quad & $3.28$  &  $ 5.44379  $  & $ 7.32457 $  &  $ 1.86129  $  & $ 266.524 $ & $ 3.63876 $ \\
Ethane ($\text{C}_2\text{H}_6$)\quad  & $4.54$  &  $ 7.96745  $  & $ 5.32158 $  &  $ 1.27174  $  & $ 311.808 $ & $ 5.85931 $ \\
\hline
\hline
\end{tabular}
\end{center}
\label{tab-values}
\end{table*}  

In Table~\ref{tab-values1} we exhibit estimates of the density $\rho(r)$, speed of sound $c_{\rm s}(r)$, pressure $P(r)$ and temperature $T(r)$ for the Ethane gas ($\text{C}_2\text{H}_6$) at different positions of the radial coordinate. 
\begin{table}[htpb!]
\caption{Estimates of the density $\rho(r)$, speed of sound $c_{\rm s}(r)$, pressure $P(r)$ and temperature $T(r)$ for the Ethane gas ($\text{C}_2\text{H}_6$) at $r/r_{\rm c}=2.0$, $r/r_{\rm c}=4.0$, $r/r_{\rm c}\rightarrow \infty$. In parentheses are indicated the units used in each quantity.} 
\begin{center}
\begin{tabular}{c c c c}
\hline\hline
 \multicolumn{1}{c}{Quantity}  &  \multicolumn{1}{c}{$r/r_{\rm c}=2.0$}  & \multicolumn{1}{c}{$r/r_{\rm c}=4.0$} & \multicolumn{1}{c}{$r/r_{\rm c}\rightarrow \infty$} \\
\hline
$\rho \,\, (\text{kg/m$^3$})$ \quad & $ 0.34448$ \quad & $ 0.94873  $ \quad & $ 1.27174 $\\
$c_{\rm s}\, \, (\text{m/s}) $ \quad &  $ 270.033 $  & $ 301.906 $ & $ 311.808 $\\
$P \,\, (\text{Pa})$ \quad &   $ 20585.2  $  & $ 70865.9$ & $ 101325$\\
$T \,\, (\text{K}) $ \quad &  $ 216.112 $  & $270.141$ & $ 288.150$\\
\hline\hline
\end{tabular}
\end{center}
\label{tab-values1}
\end{table}  

\begin{figure*}[htpb!]
   \includegraphics[width=0.5\textwidth]{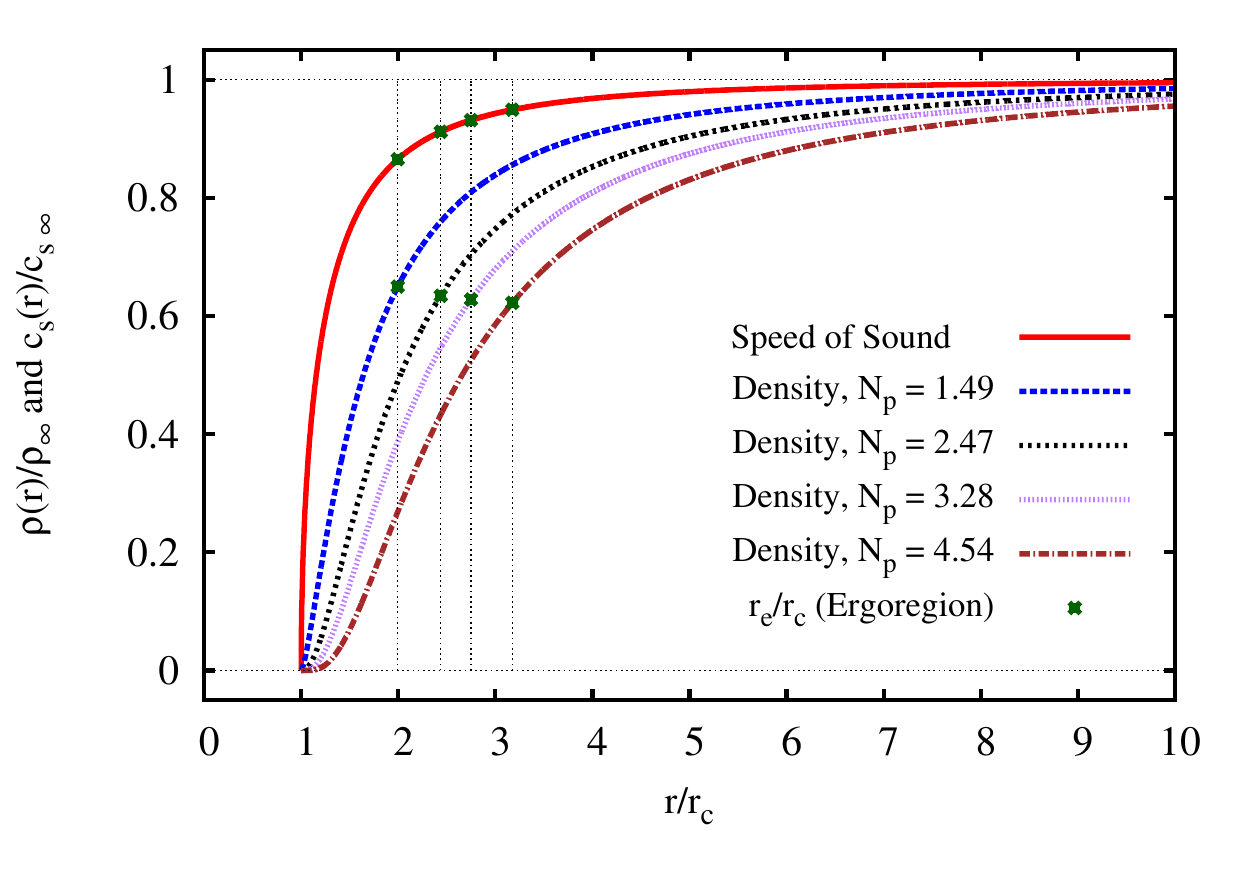}\includegraphics[width=0.5\textwidth]{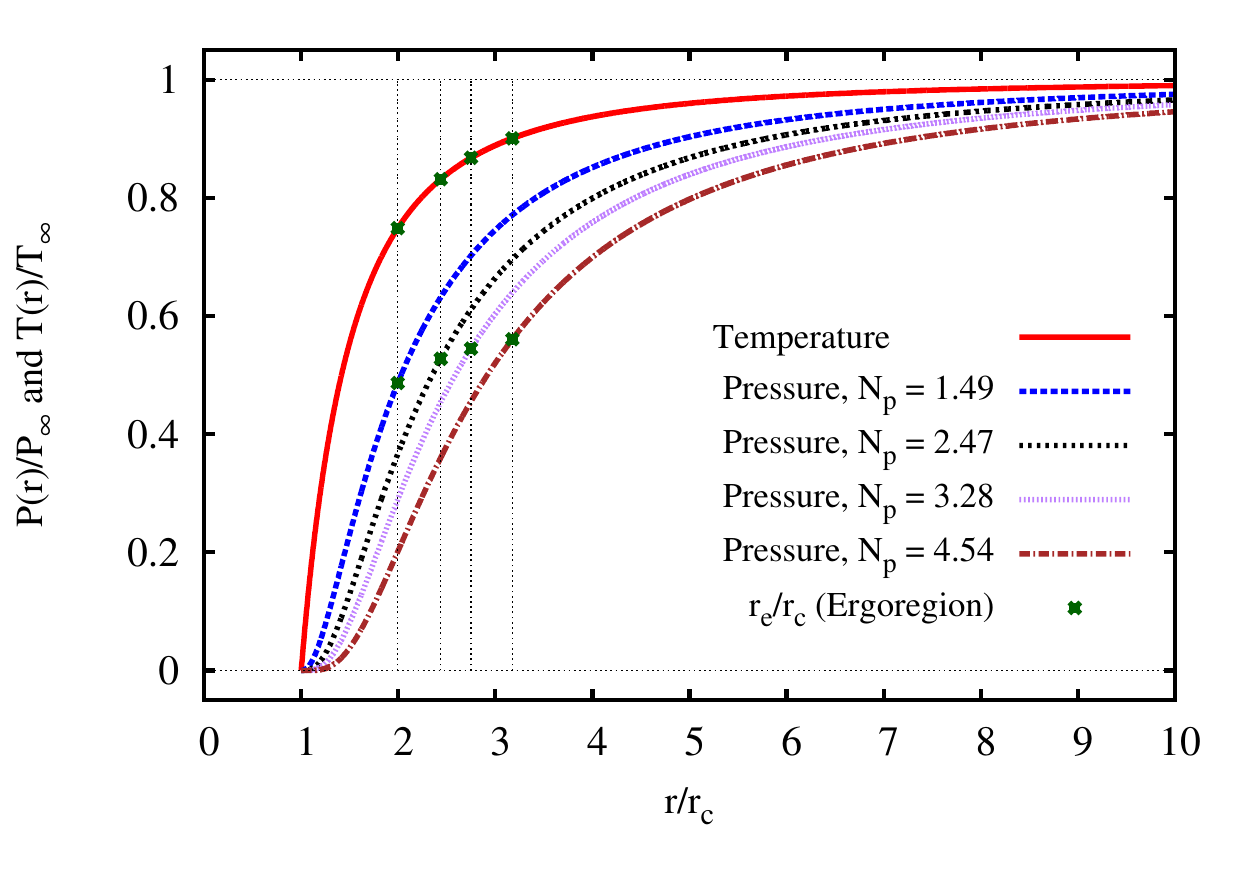}
  \caption{(Left plots) Fluid density and speed of sound as functions of $r$, for polytropic index $N_{\rm p} = 1.49$, $N_{\rm p}=2.47$, $N_{\rm p}= 3.28$ and $N_{\rm p} = 4.54$. 
  (Right plots) Pressure and temperature as functions of $r$, for the same choices of the polytropic index.}
   \label{fig-Density_Speed_Pressure_Temperature}
  \end{figure*}

In Fig.~\ref{fig-Density_Speed_Pressure_Temperature} we plot, respectively, fluid density $\rho$, speed of sound $c_{\rm s}$, pressure $P$ and temperature $T$, as functions of $r$ and for different values of polytropic index $N_{\rm p}$. 
  
\section{Perturbations of a polytropic hydrodynamic vortex}
\label{sec-Perturbations}
Linear perturbations propagating in acoustic spacetimes are described by the Klein-Gordon equation, namely
\beq
 \Box \Phi = \frac{1}{\sqrt{|g|}} \partial_a \left( \sqrt{|g|} g^{a b} \partial_b \Phi \right) = 0, 
\label{kg}
\eeq
where $g^{a  b}$ is the contravariant \emph{effective metric}, given by
\beqn
 g^{a b} =-\frac{1}{\rho c_{\rm s}}\left(
\begin{array}{cccc}
1 & 0 & C/r^2 & 0\\
\nn\\
0 & -c_{\rm s}^2 & 0 & 0\\
\nn\\
C/r^2 & 0 & (C^2/r^2-c_{\rm s}^2)/r^2 & 0\\
\nn\\
0 & 0 & 0  &-c_{\rm s}^2
\end{array}
\right)
\label{contra_metric}
\eeqn
and $g \equiv \det(g_{a  b}) = -\rho^4r^2/c_{\rm s}^2$.

The spacetime of the polytropic hydrodynamic vortex is cylindrically symmetric,
so that we may decompose the field $\Phi$ in terms of azimuthal harmonics, namely
\beq
\Phi(t,r,\theta, z) = \frac{1}{\sqrt{r}} \sum_{m=-\infty}^{\infty} \psi_m(t, r) \exp\left( {i m \theta}\right), \label{modesum}
\eeq
where $m$ is an integer number that is related with the angular momentum of the perturbation. The coordinate $z$ is trivial, and the spacetime under study is effectively three-dimensional.

Substituting the covariant metric $g^{a b}$ and Eq.~(\ref{modesum}) into Eq.~(\ref{kg}), we obtain 
\beqn
&&\left[ -\frac{1}{c_{\rm s}^2}\left(\frac{\partial}{\partial t}+\frac{iCm}{r^2}\right)^2+\frac{\partial^2}{\partial r^2}+ \frac{1}{\rho}\frac{d\rho}{dr}\frac{\partial}{\partial r}\right.\nn\\
&&\left.-\frac{1}{r^2}\left(m^2-\frac{1}{4} \right)-\frac{1}{2r\rho}\frac{d\rho}{dr}   \right] \psi_m(t,r)=0.
\label{waveeq0}
\eeqn

In the frequency domain, we assume single-frequency modes $\psi_m(t, r)$ by using the following ansatz
\beq
\psi_m(t, r) = u_{\omega m}(r)\exp\left({-i\omega t}\right).
\label{modal}
\eeq
Substituting Eq.~(\ref{modal}) into Eq.~(\ref{waveeq0}), we obtain
\beqn
&&\left[\frac{d^2}{dr^2}+ \frac{1}{\rho}\frac{d\rho}{dr}\frac{d}{dr}+\frac{1}{c_{\rm s}^2}\left(\omega-\frac{Cm}{r^2} \right)^2 \right.\nn\\
&&\left.-\frac{1}{r^2}\left(m^2-\frac{1}{4} \right)-\frac{1}{2r\rho}\frac{d\rho}{dr}\right]u_{\omega m}(r)=0.
\label{radial0}
\eeqn

Furthermore, we may use Eqs.~\eqref{density},~\eqref{critical},~\eqref{speed3}, and~\eqref{csinf} into Eq.~\eqref{waveeq0}, to obtain the following partial differential equation
\beqn
&&\left[-\left(x^2\frac{\partial}{\partial \tau}+\frac{i\sqrt{2N_{\rm p}}Cm}{|C|}\right)^2 +x^2\left(x^2-1 \right)\frac{\partial^2}{\partial x^2}+ 2x N_{\rm p}\frac{\partial}{\partial x} \right.\nn\\
&&\left.-\left(x^{2}-1 \right)\left(m^2-\frac{1}{4} \right)-N_{\rm p}\right] \psi_m(\tau,x)=0,
\label{waveeq}
\eeqn
where we defined a dimensionless time coordinate $\tau$, given by 
\beq
\tau \equiv tc_{\rm s\infty}/r_{\rm c},
\eeq
and a dimensionless radial coordinate $x$, namely
\beq
x \equiv r/r_{\rm c}.
\eeq

Equivalently, using Eqs.~\eqref{density},~\eqref{critical},~\eqref{speed3}, and~\eqref{csinf} into Eq.~\eqref{radial0}, we obtain the following ordinary differential equation for single-frequency modes $u_{\varpi m}(x)$
\beqn
&&\left[x^{2}\left(x^{2}-1 \right)\frac{d^2}{dx^2}+ 2x N_{\rm p}\frac{d}{dx}+\left(x^{2}\varpi-\frac{\sqrt{2N_{\rm p}}Cm}{|C|} \right)^2 \right.\nn\\
&&\left.-\left(x^{2}-1 \right)\left(m^2-\frac{1}{4} \right)-N_{\rm p}\right]u_{\varpi m}(x)=0,
\label{radial}
 \eeqn 
where we defined a dimensionless frequency $\varpi$, given by  
\beq
\varpi \equiv r_{\rm c}\omega/c_{\rm s\infty}.
\label{varpi}
\eeq
The ordinary differential equation given by Eq.~\eqref{radial} has regular singular points at the origin $x=0$, at the critical radius $x = x_c \equiv 1$, and at $x=-1$, and an irregular singular point at infinity ($x\rightarrow \infty$).

From Eq.~(\ref{radial}), it is straightforward to see that the polytropic hydrodynamic vortex has the following symmetries in the frequency $\varpi$, related to azimuthal number $m$ and circulation $C$
\beq
\varpi (m,C) = -\varpi^\ast (-m,C) = -\varpi^\ast (m,-C) = \varpi (-m,-C), \nn
\eeq
where ``${}^\ast$'' denotes complex conjugation. Henceforth, taking into account these symmetries, we assume without loss of generality that $m > 0$ and $C > 0$. 
  
\subsection{Boundary and initial conditions}
\label{sec-Boundary}
We consider two different boundary conditions at $x_{\rm min} \equiv r_{\rm min}/r_{\rm c}$, adopting essentially the same physically acceptable boundary conditions proposed in Ref.~\cite{Oliveira:2014oja} (i.e., we consider a cylinder with radius $r=r_{\rm min}$ made of a certain material with acoustic impedance $Z$), namely: (i) {\rm BC\,I}, a boundary condition of Dirichlet type which mimics low-$Z$ materials, and
(ii) {\rm BC\,II}, a boundary condition of Neumann type which mimics high-$Z$ materials~\cite{Lax:1948}.

The boundary condition {\rm BC\,I} may be defined as
\beq
\left[ u_{\varpi m}(x)\right]_{x = x_{\rm min}} =0\,.\label{BCI}
\eeq
The boundary condition {\rm BC\,II} is given by (cf. Eq.~\eqref{modesum})
\beq
\left[\frac{d}{dx}\left( \frac{u_{\varpi m}(x)}{\sqrt{x}}\right) \right]_{x=x_{\rm min}}=0\,.\label{BCII}
\eeq
Both these boundary conditions are realistic and 
whether {\rm BC\,I} or {\rm BC\,II} hold in practice 
depends on how the experimental apparatus is implemented in the laboratory.

At large radial distances we require outgoing Sommerfeld or causal boundary conditions, which in the frequency domain, and given our choice of the Fourier transform, amount to
\beq
u_{\varpi m}\left(x\to \infty \right) \sim \exp\left( {i\varpi x}\right)\,.\label{BC2}
\eeq

As initial condition to time-domain analysis, we use the following Gaussian package
\beq
\left[ \psi_m(\tau, x)\right]_{\tau = 0} = \left(x-x_{\rm min}\right)^2\exp\left( \frac{-\left(x -x_0\right)^2}{2 \sigma^2}\right),
\label{Gauss}
\eeq
and its first order derivative with respect to the time
\beq
\left[ \frac{\partial \psi_m }{\partial \tau} (\tau, x)\right]_{\tau = 0} = 0,
\eeq
where $x_0$ is the position of the center of the peak of the Gaussian function (middle point), and $\sigma$ sets the width of the Gaussian function. Note that the Gaussian package, given by Eq.~\eqref{Gauss}, satisfies both boundary conditions {\rm BC\,I} and {\rm BC\,II}.

Our main task consists therefore in studying solutions of Eqs.~\eqref{waveeq} and~\eqref{radial}, subjected to these boundary and initial conditions.

\section{Numerical methods} 
\label{sec-Numerical}

\subsection{Method of lines (MOL)}
\label{sec-Lines}
We may determine the time-domain profiles associated with the QNMs of the polytropic hydrodynamic vortex applying a numerical method to solve the partial differential equation (\ref{waveeq}). We study the evolution of a Gaussian disturbance in the time domain, using the method of lines (MOL) as a numerical simulation \cite{Rinne, Witek:2012tr}. The MOL involves a second-order spatial coordinate discretization and fourth-order Runge-Kutta method to advance in time~\cite{Dolan:2011ti, Dolan:2012yt, Butcher}.
       
To apply the MOL in Eq.~(\ref{waveeq}), we discretize the radial coordinate $x \rightarrow x_{j} = x_{\rm min}+jh$ (for a range $x_{\rm min} \leq x \leq x_{\rm max}$), the wave function $\psi_m(\tau, x) \rightarrow \psi_j$, first-order spatial derivative $\dfrac{\partial}{\partial x} \psi_m(\tau, x) \rightarrow \xi_j$, with
\beq
\xi_j=\frac{1}{2h}\left(\psi_{j+1}-\psi_{j-1}\right) +{\cal{O}}(h^2);
\eeq
 and second-order spatial derivative $\frac{\partial^2 }{\partial x^{2}} \psi_m(\tau, x) \rightarrow \chi_j$, namely,
\beq
\chi_j=\frac{1}{h^2}\left(\psi_{j+1}-2\psi_{j}+\psi_{j-1}\right) +{\cal{O}}(h^2),
\eeq
where $h$ is the grid spacing \cite{Rinne, Butcher}. 
 
Employing the discretizations and transforming the spatial second-order derivative using the definition $\zeta_{j} \equiv \dfrac{d \psi_{j}}{d\tau}$, we may obtain from Eq.~\eqref{waveeq} a set of two first-order differential equations. Then, we apply the fourth-order Runge-Kutta method in each first-order differential equation to evolve the Gaussian perturbation (\ref{Gauss}) in time domain (for a range $0 \leq \tau \leq \tau_{\rm max}$) \cite{Dolan:2011ti}. 

\subsection{Direct integration (DI) method} 
\label{sec-Direct}

In the frequency-domain, the differential equation~(\ref{radial}) subjected to the boundary conditions described in Section~\ref{sec-Boundary} is an eigenvalue problem for the frequency $\omega$. Frequencies that satisfy the eigenvalue problem are called QNM frequencies.
To investigate superradiant instabilities, we compute the QNM frequencies for the polytropic hydrodynamic vortex. QNMs are generically modes with complex frequencies. The time-dependence of the fluctuations, expressed in Eq.~(\ref{modal}), implies that if the imaginary part of the QNM frequencies is negative ($\text{Im}[\omega] < 0$) the spacetime is stable and the perturbations vanish for late times as $\psi \sim \exp\left({-|\text{Im}[\omega]| t}\right)$. In contrast, if the imaginary part of the QNM frequencies is positive ($\text{Im}[\omega] > 0$), the perturbations are amplified as $\psi \sim \exp\left({\text{Im}[\omega]t}\right)$ and the spacetime is unstable.

We may solve directly the differential equation~(\ref{radial}) using a direct integration (DI) method to obtain the QNM frequencies. 
The DI is based on the shooting method and numerical root-finding to obtain frequencies in the complex domain \cite{Dolan:2010zza}. We write the outgoing solution at infinity as a generalized power series 
\beq 
u_{\varpi m}\left(x\to \infty \right) \sim \exp\left( {i\varpi x}\right) \sum_{i=0}{\frac{b_i}{ x^i}}\,.
\label{serie2}
\eeq
The series~(\ref{serie2}) and its first-order derivative are then used as boundary conditions to directly integrate Eq.~(\ref{radial}) 
inwards for a range $x_{\rm max} \geq x \geq x_{\rm min}$. 

The QNM frequencies are the roots of this integration procedure that satisfy the appropriate {\rm BC\,I} or {\rm BC\,II} at $x=x_{\rm min}$, i.e., Eqs.~\eqref{BCI} and \eqref{BCII}, respectively. To find these roots, we use standard root-finding algorithms such as Newton's method.

\subsection{Continued fraction (CF) method} 
\label{sec-Continued}
An alternative to directly integrating Eq.~(\ref{radial}), consists in expressing the problem
as a continued fraction (CF) to find the QNM frequencies \cite{Leaver:1985ax}.
We may define the Frobenius-like series in the neighborhood of $x = x_{\rm min}$, namely
\beq
u_{\varpi m}(x) = \exp\left( {i\varpi x}\right) \sum_{n=0} a_n \left( 1 -\frac{ x_{\rm min}}{x} \right)^{n}\,.   \label{cf-ansatz-vt}
\eeq
Substituting Eq.~(\ref{cf-ansatz-vt}) into Eq.~(\ref{radial}), we find the following five-term recurrence relation:
\beqs \label{reco-vt}
&&\alpha_0a_2+\beta_0a_1 +\gamma_0a_0= 0, \\
&&\alpha_1a_3+\beta_1a_2+\gamma_1a_1 +\delta_1a_0 = 0,  \\
&&\alpha_n a_{n+2}+\beta_na_{n+1}+\gamma_n a_{n}+\delta_{n}a_{n-1}+\epsilon_{n}a_{n-2} = 0, \nn\\ &&\text{for} \hspace{0.1cm} n \geq 2,
\eeqs
where the recurrence coefficients are given by
\beqs\label{coeff-reco}
\alpha_n &=& 4 (1+n) (2+n) \left(-1+x_{\rm min}^2\right),\\
\beta_n &=& 8 (1+n) \left[N_{\rm p}-n \left(-2+x_{\rm min}^2\right)\right.\nn\\
&+&\left.i \left(-1+x_{\rm min}^2\right) (i+x_{\rm min} \varpi )\right],  \\
\gamma_n &=& 4 \left[m^2+2 m^2 N_{\rm p}-6 n N_{\rm p}+2 i (2 n+N_{\rm p}) x_{\rm min} \varpi\right.\nn\\
&+&\left.x_{\rm min}^2 \left(-m^2+n+n^2-2 \sqrt{2} m \sqrt{N_{\rm p}} \varpi +\varpi ^2\right)\right]\nn\\
&-&1-24 n^2-4 N_{\rm p}+x_{\rm min}^2, \\
\delta_{n} &=& 2 \left[5-12 n+8 n^2-8 N_{\rm p}+12 n N_{\rm p}-4 m^2 (1+2 N_{\rm p})\right.\nn\\
&-&\left.4 i (-1+n+N_{\rm p}) x_{\rm min} \varpi \right],\\
\epsilon_{n}&=& -(-3+2 n) (-3+2 n+4 N_{\rm p})+m^2 (4+8 N_{\rm p}) .
\eeqs

Using a double Gaussian elimination (cf. Refs.~\cite{Oliveira:2014oja, Onozawa:1995vu}) from the five-term recurrence relation~\eqref{reco-vt} we may write the following three-term recurrence relation:
\beqn
\alpha_n a_{n+2}+\beta_na_{n+1}+\gamma_n a_{n} = 0, \hspace{0.5cm} \text{for} \hspace{0.1cm} n \geq 0.
\label{reco-vt-3}
\eeqn
The recurrence coefficients $ \alpha_n, \beta_n $ and $ \gamma_n $ are complex functions that depend on the frequency $ \varpi $, the azimuthal number $ m $, the polytropic index $ N_p $ and $ x_{\rm min} $. 

For {\rm BC\,I}, considering $a_0=0$ in Eq.~(\ref{reco-vt-3}), we obtain the following continued-fraction 
\beqn
\beta_0-\dfrac{\alpha_0\gamma_{1}}{\beta_{1}-\dfrac{\alpha_{1}\gamma_{2}}{\beta_{2}-\dfrac{\alpha_{2}\gamma_{3}}{\beta_{3}-...}}} = 0. 
\label{highordern-vt}
\eeqn

From Eq.~\eqref{cf-ansatz-vt} and considering $n \rightarrow n-1$ in Eq.~(\ref{reco-vt-3}) we find the following relation for {\rm BC\,II} 
\beq
1 -2i\varpi x_{\rm min} +\dfrac{2\gamma_1}{\beta_1-\dfrac{\alpha_1\gamma_2}{\beta_2-\dfrac{\alpha_2\gamma_3}{\beta_3-...}}}=0.
\label{frac_bcii}
\eeq

Eqs.~(\ref{highordern-vt}) and~(\ref{frac_bcii}) can be solved with standard root-finding algorithms such as Newton's method.

\section{Results} 
\label{sec-Results}
Concerning the radial position $r_{\rm min}$ where we impose boundary conditions {\rm BC\,I} and {\rm BC\,II}, the QNMs of the polytropic hydrodynamic vortex may be separated in two categories:\\
{\bf (i)} QNMs for $r_{\rm min}$ located outside of the outer boundary of the ergoregion $r_{\rm e}$, which in dimensionless radial coordinate can be written as $x_{\rm min} \geq \sqrt{2N_{\rm p}+1}$ [see Eq.~\eqref{re_rc}]. \\
{\bf (ii)} QNMs for $r_{\rm min}$ located inside of the outer boundary of the ergoregion $r_{\rm e}$,  $1<x_{\rm min} < \sqrt{2N_{\rm p}+1}$.

For $r_{\rm min}$ located outside the outer boundary $r_{\rm e}$ of the ergoregion, the polytropic hydrodynamic vortex admits only stable QNM frequencies; 	
whereas for $r_{\rm min}$ located inside $r_{\rm e}$, the polytropic hydrodynamic vortex admits both stable and unstable QNM frequencies~\cite{Oliveira:2014oja}. Here we are especially interested in computing the QNM frequencies for $r_{\rm min}$ located inside $r_{\rm e}$, which exhibit the ergoregion instabilities associated with this system.

We follow standard conventions of ordering the QNM frequencies $\varpi$ (in dimensionless units) by their imaginary part~\cite{Berti:2009kk}.
The fundamental mode is the one with largest imaginary component ${\rm Im}[\varpi]$.
Thus, if the mode is unstable (${\rm Im}[\varpi]>0$), the fundamental mode corresponds to the smallest instability
timescale, and for stable (${\rm Im}[\varpi]<0$) modes it corresponds to the longest-lived mode.

\subsection{Boundary conditions imposed outside of the outer boundary of the ergoregion} 
\label{sec-outside}
We have computed QNM frequencies for $r_{\rm min}$ located outside of the outer boundary $r_{\rm e}$ of the ergoregion, checking the agreement between three methods for different values of the polytropic index $N_{\rm p}$. Examples are shown in Table~\ref{tab-freq_1}. 

In Table~\ref{tab-freq_1} we exhibit the estimates of the QNM frequencies $\varpi$ obtained via MOL, DI and CF methods for different values polytropic index $N_{\rm p}$. We have obtained excellent agreement between three methods.
\begin{table}[htpb!]
\caption{QNM frequencies $\varpi$ for azimuthal number $m=3$, and $x_{\rm min}=4.0$, obtained numerically from estimates via MOL, DI and CF methods, for polytropic index $N_{\rm p} = 1.49$, $N_{\rm p}=2.47$, $N_{\rm p}= 3.28$ and $N_{\rm p} = 4.54$ using boundary conditions {\rm BC\,I} and {\rm BC\,II}. For MOL, we compute QNM frequencies from time-domain profiles $|Re(\psi_m(\tau, x))|$ extracted at $x = 12.0$, via Eq.~\eqref{waveeq}. Here we use a grid spacing $h=1/1000$, width of Gaussian function $\sigma=0.25$, and middle point of Gaussian function $x_0=6.0$. Note that the QNM frequency $\varpi$ is dimensionless, for conversion into dimensional frequency $\omega$, one should multiply by $c_{\rm s \infty} /r_{\rm c}$ (cf. the data in Table~\ref{tab-values}).}
\begin{center}
\begin{tabular}{c c c c c c}
\hline\hline
 \multicolumn{1}{c}{} & \multicolumn{1}{c}{} & \multicolumn{2}{c}{{\rm BC\,I}} & \multicolumn{2}{c}{{\rm BC\,II}} \\
 \multicolumn{1}{c}{$N_{\rm p}$} & \multicolumn{1}{c}{Method} & \multicolumn{1}{c}{$\text{Re}(\varpi) $} & \multicolumn{1}{c}{$\text{Im}(\varpi) $}& \multicolumn{1}{c}{$\text{Re}(\varpi) $}& \multicolumn{1}{c}{$\text{Im}(\varpi) $} \\
\hline
\multirow{3}{*}{$1.49$} & MOL & $ -0.277610 $& $-0.230845 $ & $ -0.373637 $& $ -0.110406  $ \\
& DI & $ -0.277939 $& $- 0.235285 $ & $  -0.374032 $& $- 0.109901 $ \\
& CF & $ -0.277323 $& $- 0.232678  $ & $ -0.374032 $& $ - 0.109901 $ \\
\hline
\multirow{3}{*}{$2.47$} & MOL & $ -0.262089 $& $-0.191755   $ & $ -0.322674 $& $-0.080349$\\
& DI & $ -0.262375 $& $ -0.192064 $ & $  -0.322619 $& $- 0.080100 $ \\
& CF & $ -0.262819 $& $ - 0.191891 $  & $ -0.322619 $& $ - 0.080100 $\\
\hline
\multirow{3}{*}{$3.28$} & MOL & $ -0.251992 $& $ -0.166754 $ & $ -0.290045 $& $ -0.061799  $\\
& DI & $ -0.252594 $& $ - 0.166545  $  & $ -0.289911 $& $- 0.061835 $ \\
& CF & $ -0.252738 $& $ - 0.166635  $  & $ -0.289911 $& $ - 0.061835  $ \\
\hline 
\multirow{3}{*}{$4.54$} & MOL &  $ -0.238283 $& $-0.136121 $ & $ -0.248886$& $-0.040595$\\
& DI & $ -0.238832 $& $ - 0.135947 $ & $ -0.248850 $& $ - 0.040702  $ \\
& CF & $ -0.238817 $& $ - 0.135995 $ & $ -0.248850 $& $ - 0.040702 $ \\
\hline\hline 
\end{tabular}
\end{center}
\label{tab-freq_1}
\end{table}

\subsection{Boundary conditions imposed inside of the outer boundary of the ergoregion} 
\label{sec-intside}
Next, we choose $r_{\rm min}$ located inside of the outer boundary $r_{\rm e}$ of the ergoregion and outside of the critical radius $r_{\rm c}$ ($1<x_{\rm min} < \sqrt{2N_{\rm p}+1}$).

In Table~\ref{tab-freq_2} we exhibit the estimates of the QNM frequencies $\varpi$ for azimuthal number $m=5$, obtained via MOL, DI and CF methods, for different values of polytropic index $N_{\rm p}$.
\begin{table}[htpb!]
\caption{QNM frequencies $\varpi$ for azimuthal number $m=5$, obtained numerically from estimates via MOL, DI and CF methods, for polytropic index $N_{\rm p}=2.47$, $N_{\rm p}=3.28$ and $N_{\rm p}=4.54$. We impose boundary conditions {\rm BC\,I} and {\rm BC\,II} at $x_{\rm min}=2.0$. For MOL, we compute QNM frequencies from time-domain profiles $|Re(\psi_m(\tau, x))|$ extracted at $x = 10.0$, via Eq.~\eqref{waveeq}. We use a grid spacing $h=1/1000$, width of Gaussian function $\sigma=0.25$, and middle point of Gaussian function $x_0=4.0$.}
\begin{center}
\begin{tabular}{c c c c}
\hline\hline
 \multicolumn{1}{c}{} & \multicolumn{1}{c}{} & \multicolumn{2}{c}{{\rm BC\,I}} \\
 \multicolumn{1}{c}{$N_{\rm p}$} & \multicolumn{1}{c}{Method} & \multicolumn{1}{c}{$\text{Re}(\varpi) $} & \multicolumn{1}{c}{$\text{Im}(\varpi) $} \\
\hline 
\multirow{3}{*}{$2.47$} & MOL & $ -0.536536 $ & $-0.024658 $ \\
& DI & $ -0.536540 $ & $- 0.024618 $ \\
 & CF &  $ -0.536540 $ & $ - 0.024618i $   \\
\hline
\multirow{3}{*}{$3.28$} & MOL & $ -0.380340 $ & $ -0.001655   $ \\
& DI & $ -0.380328 $ & $ - 0.001675  $ \\
 & CF &  $ -0.380328 $ & $ - 0.001675 $ \\
\hline
\multirow{2}{*}{$4.54$} & DI & $ -0.123292 $ & $ -5.329260\times 10^{-9} $ \\
 & CF &  $ -0.123292 $ & $ -5.328963 \times 10^{-9}  $ \\
\hline
\end{tabular}
\begin{tabular}{c c c c }
\hline
 \multicolumn{1}{c}{} & \multicolumn{1}{c}{} & \multicolumn{2}{c}{{\rm BC\,II}} \\
 \multicolumn{1}{c}{$N_{\rm p}$} & \multicolumn{1}{c}{Method} & \multicolumn{1}{c}{$\text{Re}(\varpi) $} & \multicolumn{1}{c}{$\text{Im}(\varpi) $} \\
\hline
\multirow{2}{*}{$2.47$} & DI & $ -0.136950 $ & $ -1.172176\times 10^{-9}  $\\
 & CF &  $ -0.136950 $ & $ - 1.172059\times 10^{-9}  $   \\
\hline
\multirow{2}{*}{$3.28$} & DI & $ +0.143251 $ & $ + 3.268528\times 10^{-11}  $ \\
 & CF &  $ +0.143251 $ & $ + 3.268469\times 10^{-11}  $ \\
\hline 
\multirow{2}{*}{$4.54$} & DI & $+0.519010 $ & $ + 7.000436\times 10^{-7} $ \\
 & CF &  $ +0.519010 $ & $ + 7.000436\times 10^{-7}  $ \\
\hline\hline
\end{tabular}
\end{center}
\label{tab-freq_2}
\end{table}  

\subsection{Ergoregion instability of the polytropic hydrodynamic vortex}
\label{sec-Ergo}

\begin{figure*}[htpb!]
   \includegraphics[width=0.5\textwidth]{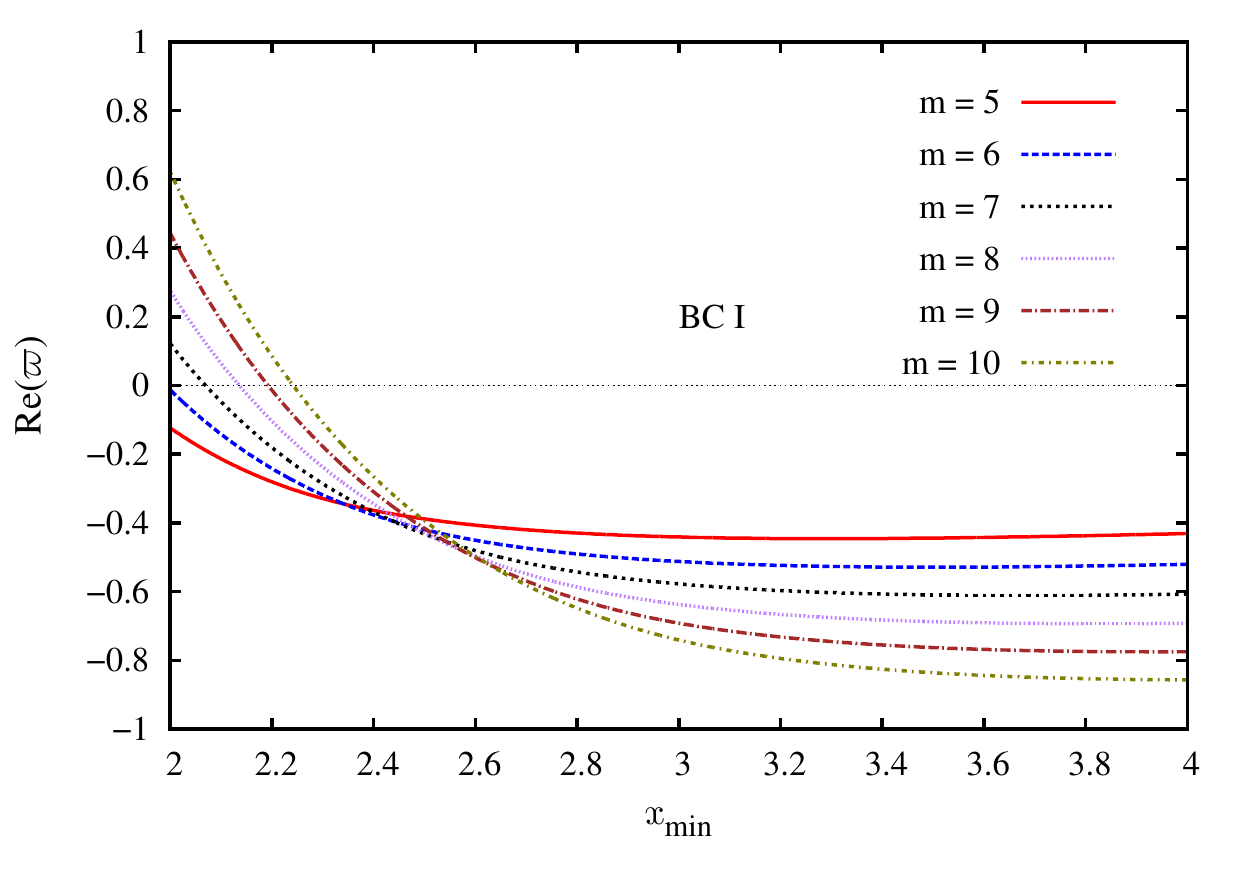}\includegraphics[width=0.5\textwidth]{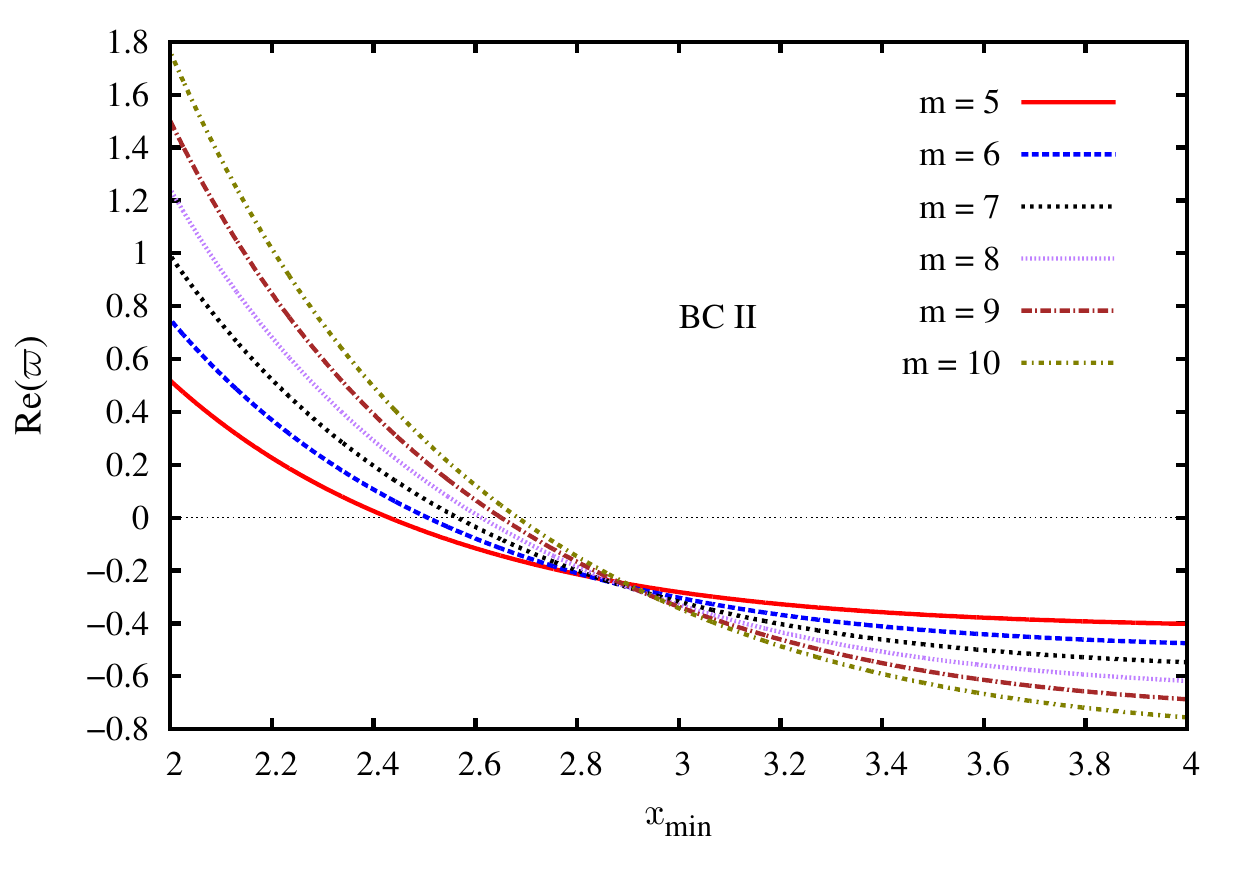}\\
   \includegraphics[width=0.5\textwidth]
   {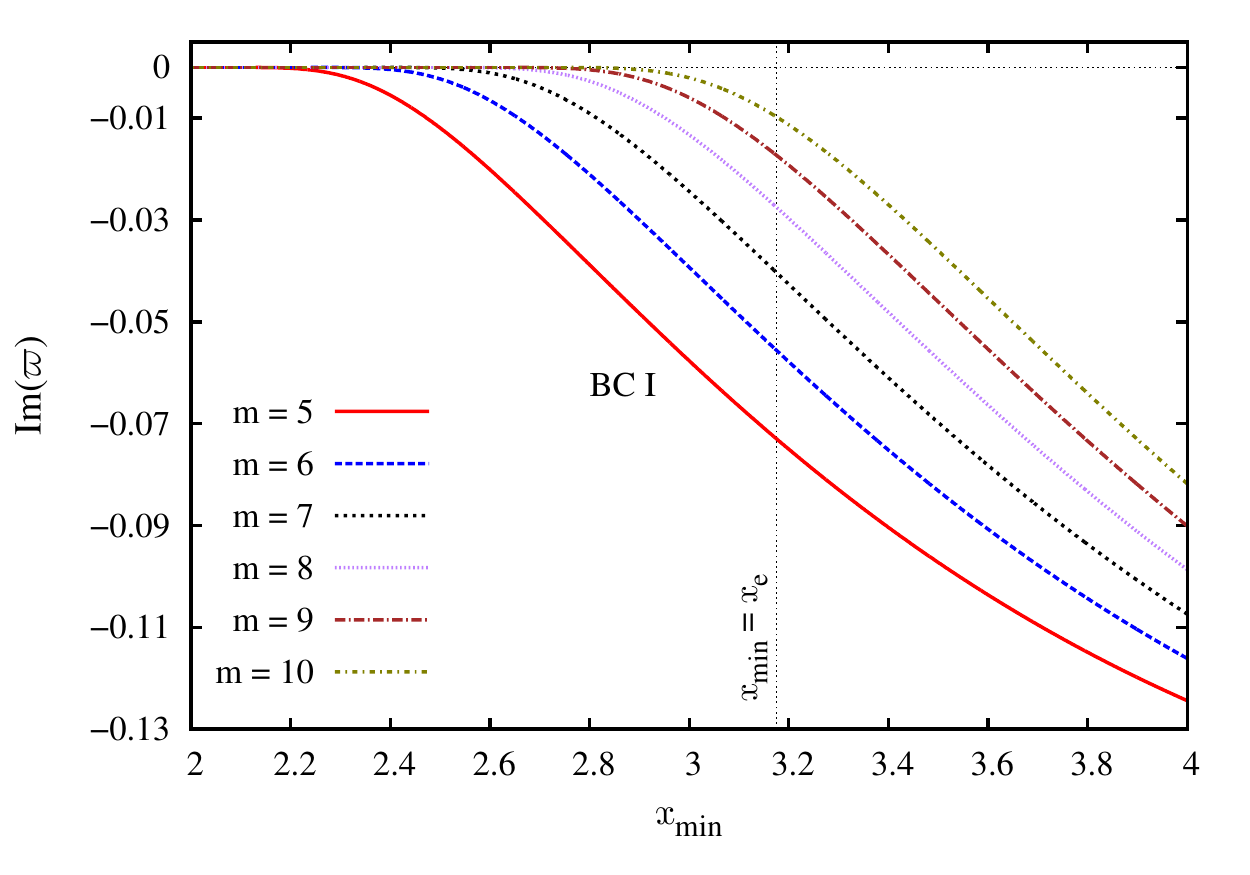}\includegraphics[width=0.5\textwidth]{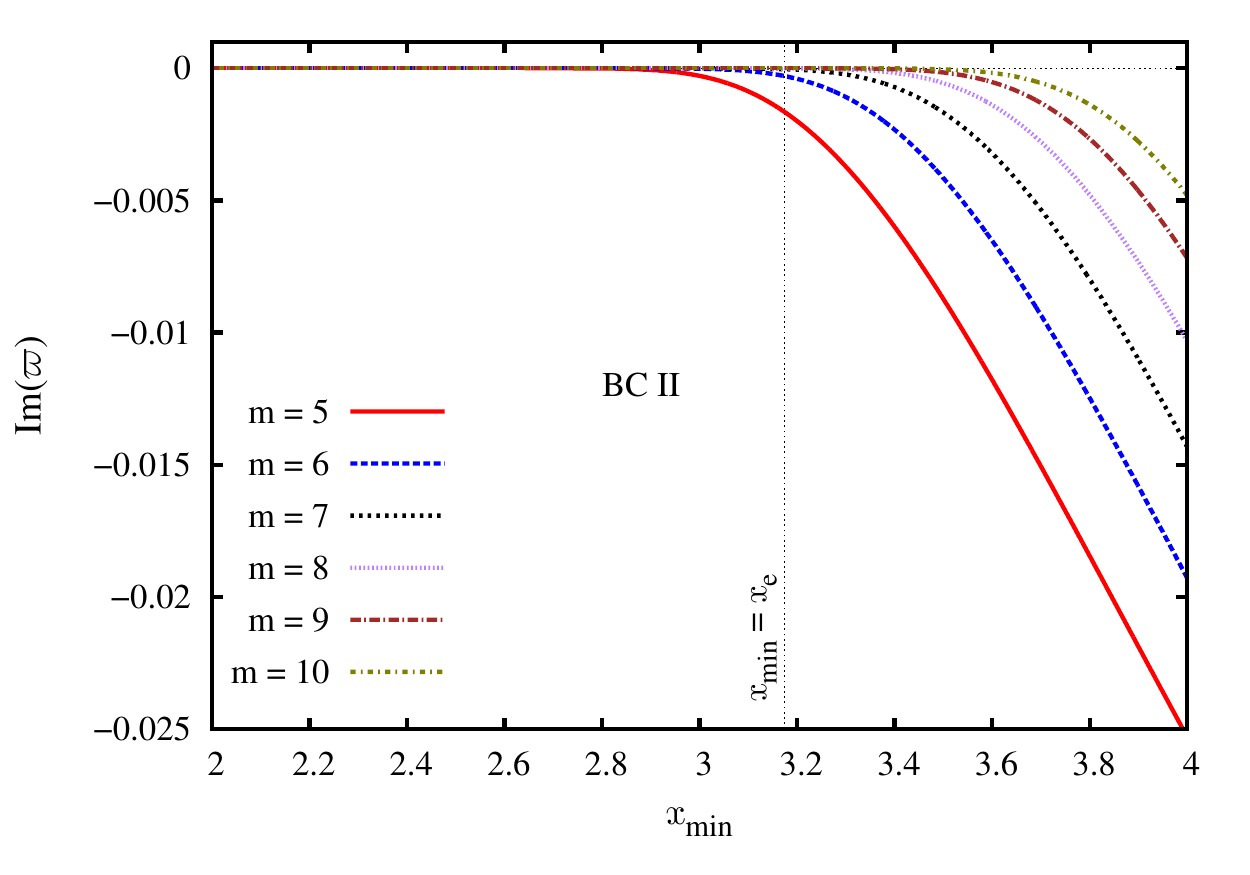}\\
   \includegraphics[width=0.5\textwidth]{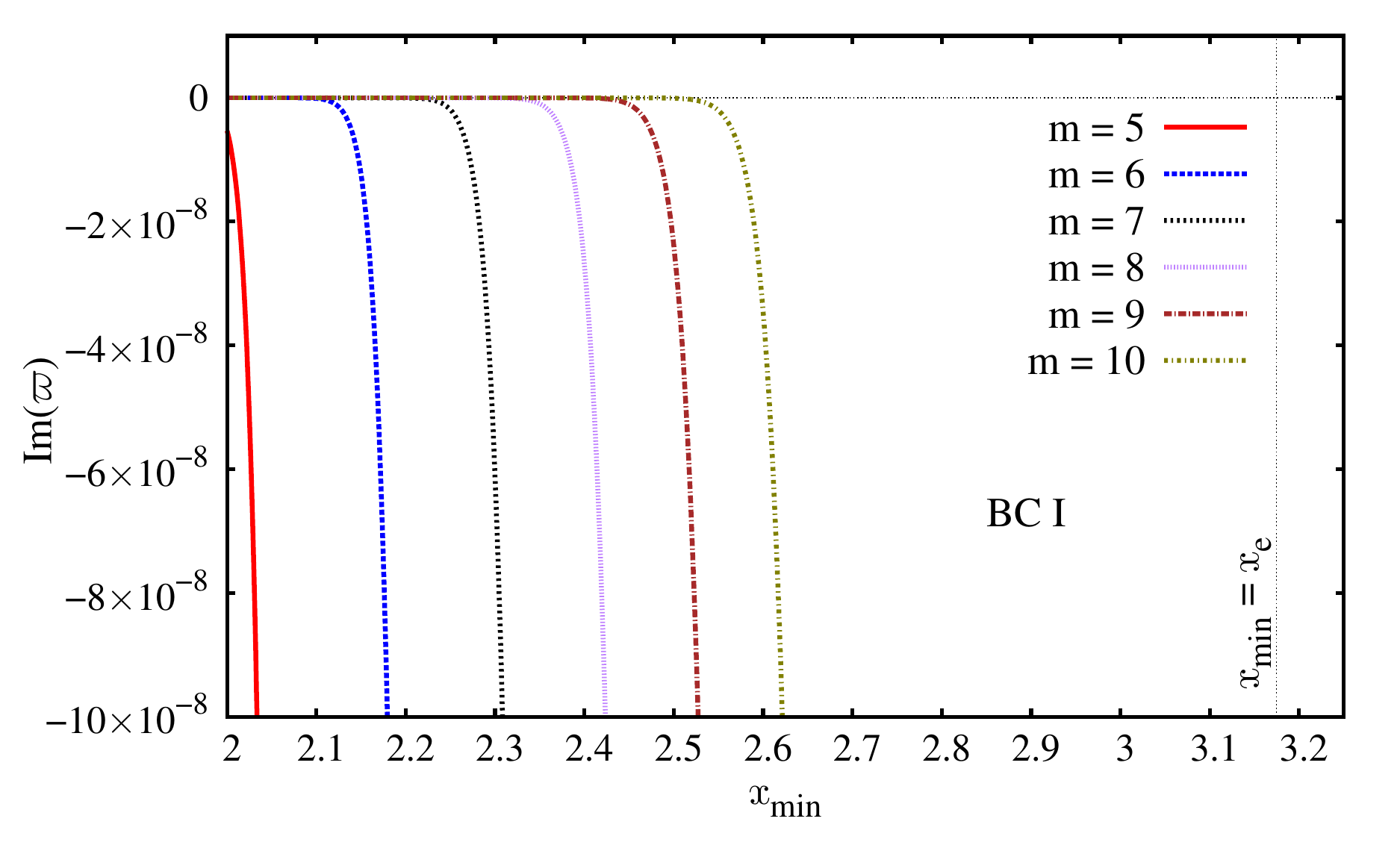}\includegraphics[width=0.5\textwidth]{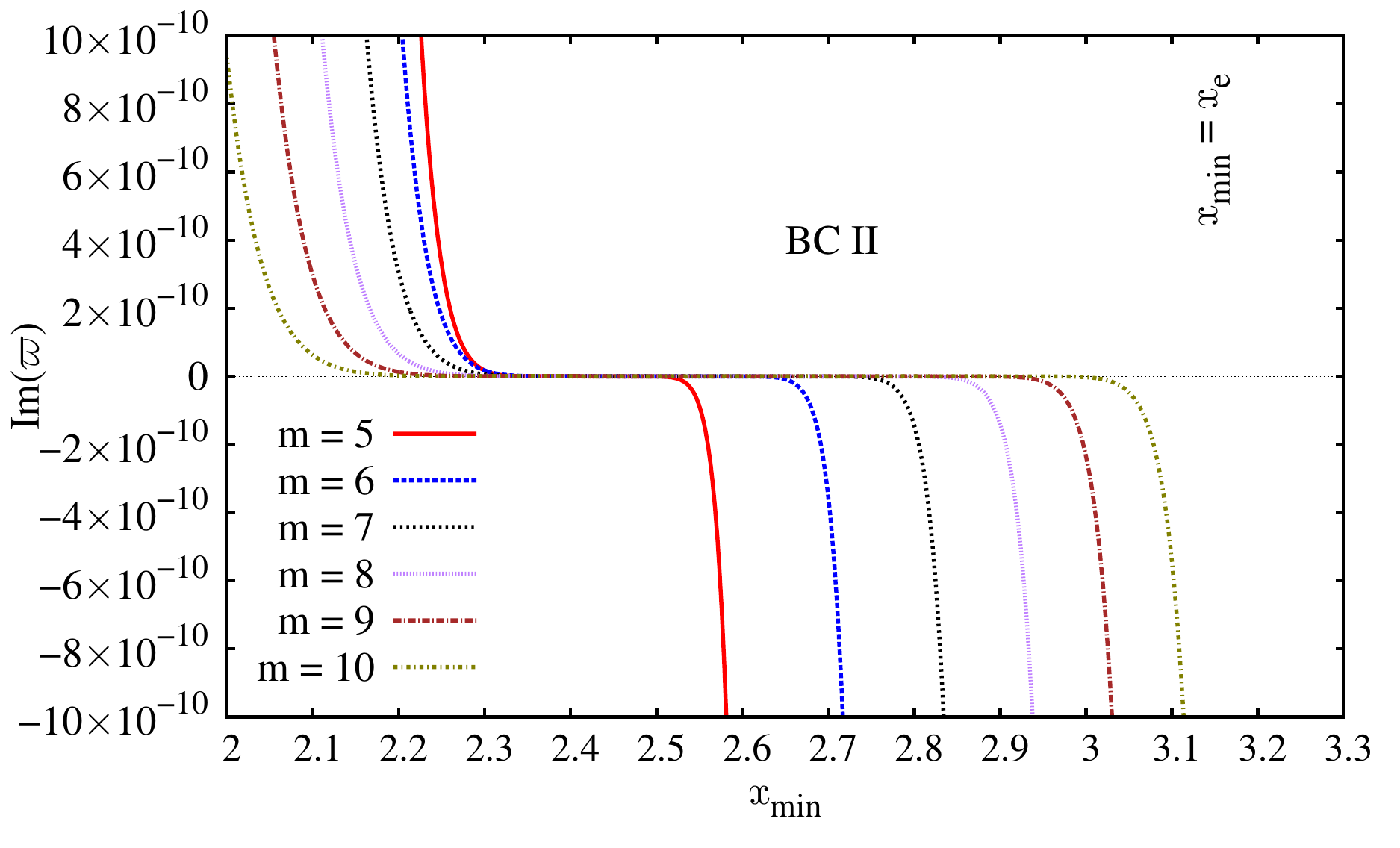}
  \caption{Real (top plots) and imaginary (middle and bottom plots) parts of the fundamental ($n = 0$) QNM frequencies $\varpi$, as a function of $x_{\rm min}$, for polytropic index $N_{\rm p}=4.54$ and different values of azimuthal numbers $m$, obtained via CF method. Here we imposed boundary conditions {\rm BC\,I} (left) and  {\rm BC\,II} (right) at $x = x_{\rm min}$. The dotted vertical lines at $x_{\rm min} = x_{\rm e}$, in the middle and bottom plots, correspond to the position of the outer boundary $r_{\rm e}$ of the ergoregion. Note that the QNM frequency $\varpi$ and radial position $x_{\rm min}$ are dimensionless. For conversion to the frequency $\omega$, one should multiply $\varpi$ by $5.85931\times 10^{5} \,\text{s}^{-1}$ and to obtain the radial position $r_{\rm min}$, one should multiply $x_{\rm min}$ by $5.32158\times 10^{-4} \,\text{m} $ (cf. the data in Table~\ref{tab-values}).}
   \label{Freq_rmin_BCI_II}
  \end{figure*}

 In Fig.~\ref{Freq_rmin_BCI_II} we plot real and imaginary parts of the fundamental ($n = 0$) QNM frequencies $\varpi$, for different values of azimuthal numbers $m$ and polytropic index $N_{\rm p}=4.54$, obtained via CF method, imposing boundary conditions {\rm BC\,I} and {\rm BC\,II} at different values of $x_{\rm min}$. Analyzing the imaginary part of the QNM frequencies $\varpi$, we conclude that, as the azimuthal number $m$ increases, the threshold between stability and instability (i.e., in the neighborhood of $\text{Im}(\varpi)=0$) increases, tending to the outer boundary  $r_{\rm e}$ of the ergoregion, which in dimensionless radial coordinate may be represented by $x_{\rm min}=x_{\rm e}$ (where $x_{\rm e}=r_{\rm e}/r_{\rm c}$). This behavior can be seen more clearly in the zooms for {\rm BC\,I} and {\rm BC\,II} exhibited in the bottom plots of Fig.~\ref{Freq_rmin_BCI_II}\,, being most prominent to boundary conditions {\rm BC\,II} than {\rm BC\,I}. 
  
As an example of a possible experimental implementation, we estimate the QNM frequencies $\omega$ for a vortex made of Ethane gas (whose polytropic constant $K_{\rm p}= 7.96745 \times 10^4\,\, \text{m}^2/\text{s}^2$ and polytropic index $N_{\rm p}=4.54$). We may obtain the QNM frequencies $\omega$ using Eq.~\eqref{varpi}, and the data exhibited in Tables~\ref{tab-values},~\ref{tab-values1} and~\ref{tab-freq_2}\,. To this particular experimental setup, we consider the circulation $C=0.5\,\, \text{m}^2/\text{s}$, subject to a perturbation with azimuthal number $m=5$ and we impose boundary conditions of Neumann type ({\rm BC\,II}) at $r_{\rm min} = 2.0 r_{\rm c}$. The real and imaginary parts of the fundamental ($n=0$) QNM frequency $\omega$ are then, respectively, $\text{Re}(\omega)= 0.304104\,\, \text{MHz}$ and $\text{Im}(\omega)= 0.410177\,\, \text{s}^{-1}$. From the imaginary part it is possible to estimate the instability timescale of this perturbation, which is $t_{\text{scale}} \equiv 1/$\text{Im}$(\omega) = 2.43797\,\,\text{s}$.

\subsection{Static (marginally-stable) resonances}
\label{sec-Static}

  Now, we investigate a critical configuration of marginal-stability of the polytropic hydrodynamic vortex, referred to as static (marginally-stable) resonances~\cite{Hod:2014hda, Marecki}, mathematically represented by $\text{Re}\left(\omega\right)=\text{Im}\left(\omega\right)=0$. (It may be seen in Fig.~\ref{Freq_rmin_BCI_II} that for specific values of $x_{\rm min}$ the QNMs frequencies goes to zero.) An analytical study of a particular case of the incompressible hydrodynamic vortex ($N_{\rm p}=0$), in its marginally-stable regime, was addressed in Ref.~\cite{Hod:2014hda}. Here we study a compressible system characterized by the polytropic equation of state~\eqref{state}.

Considering $\omega = 0$, we may obtain the following analytical solution for the ordinary differential equation~\eqref{radial}:
\beqn
u_{0m}\left(x \right)=a_1a^{+} f^-(x)+ a_2a^{-} f^+(x),
\label{sol}
\eeqn
where $a_1$ and $a_2$ are integration constants,~$a^{\pm}=i^{7/2\mp m} x^{1/2\pm m}$,
\beqn
f^\pm(x) &=& {}_2\text{F}_1\left(\frac{N_{\rm p}-b\pm m}{2},\frac{N_{\rm p}+b \pm m}{2}; 1 \pm m; x^{-2}\right),\nn\\
\eeqn
with $_2F_1\left[\alpha,\ \beta,\ \gamma, \ y\right]$ being the hypergeometric function~\cite{Grad} and $b = \sqrt{N_{\rm p}^2+m^2 (1+2N_{\rm p})}$.

We assume that $a_1=0$, so that the physically acceptable solution~\eqref{sol} is finite at $x\rightarrow \infty$. It follows that
\beqn
u_{0m}\left(x \right)=a_2 a^{-} f^+(x).
\label{sol1}
\eeqn

Applying the boundary conditions, {\rm BC\,I} [given by Eq.~\eqref{BCI}] and {\rm BC\,II} [given by Eq.~\eqref{BCII}] in the solution given by Eq.~\eqref{sol1}, we may write the following equations 
\beqn
\left[f^+(x)\right]_{x=x_{\rm min}^{(n)}} =0 \hspace{0.5cm} (\text{for {\rm BC\,I}}),
\label{root_BCI}\\
\nn\\
\left[\frac{mf^+}{x}-\frac{df^+}{dx}\right]_{x=x_{\rm min}^{(n)}}=0 \hspace{0.5cm} (\text{for {\rm BC\,II}}),
\label{root_BCII}
\eeqn
respectively, where $x_{\rm min}^{(n)}$ is the nth positive root of Eqs.~\eqref{root_BCI} (for {\rm BC\,I}) and~\eqref{root_BCII} (for {\rm BC\,II}) and describe the static (marginally-stable) resonances of the polytropic hydrodynamic vortex. To find these roots, we use a standard root-finding algorithms such as Newton's method. 

In Table~\ref{tab-roots_1} we exhibit the estimates of $x_{\rm min}^{(0)}$ (where $n=0$ denotes the fundamental QNM frequencies) for polytropic index $N_{\rm p}=4.54$ and for different values azimuthal number $m$, obtained numerically. Note that as the azimuthal number $m$ increases, the values of $x_{\rm min}^{(0)}$ tend to the values of the outer boundary $x_{\rm e}$ of the ergoregion (being most prominent to $m=1000$ and to boundary conditions {\rm BC\,II} than {\rm BC\,I}). The results exhibited in Table~\ref{tab-roots_1} are in excellent agreement with the data extracted from Fig.~\ref{Freq_rmin_BCI_II}\,, this may be denoted when we compare with the results for $x_{\rm min}^{(0)}$ in parentheses, which are obtained via CF method (some are exhibited in Fig.~\ref{Freq_rmin_BCI_II}\,), with the results obtained from Eqs.~\eqref{root_BCI} and~\eqref{root_BCII}.
\begin{table}[htpb!]
\caption{Estimates of $x_{\rm min}^{(0)}$ [the 0th root of Eqs.~\eqref{root_BCI} (for {\rm BC\,I}) and~\eqref{root_BCII} (for {\rm BC\,II})], obtained numerically using a root-finding algorithm, for polytropic index $N_{\rm p} = 4.54$, and for azimuthal number $m=6,\, 7,\, 8,\, 9,\, 10,\, 100,\, 1000$, using boundary conditions {\rm BC\,I} and {\rm BC\,II}. The results exhibited in parentheses were obtained via CF method (some are exhibited in Fig.~\ref{Freq_rmin_BCI_II}\,).}
\begin{center}
\begin{tabular}{c c c }
  \hline\hline
  \multicolumn{1}{c}{$m$} & \multicolumn{1}{c}{$x_{\rm min}^{(0)}$ ({\rm BC\,I})} & \multicolumn{1}{c}{$x_{\rm min}^{(0)}$ ({\rm BC\,II})}\\
  \hline
  $ 6 $     & $ 1.991469762231 \, (1.991) $ \quad\quad & $ 2.506392063437  \, (2.506) $ \\
  $ 7 $     & $ 2.069839310556 \, (2.070) $ \quad\quad & $ 2.565087771362  \, (2.565) $ \\
  $ 8 $     & $ 2.135952625074 \, (2.136) $ \quad\quad & $ 2.612175436164  \, (2.612) $ \\
  $ 9 $     & $ 2.192645873342 \, (2.193) $ \quad\quad & $ 2.650971380111  \, (2.651) $ \\
  $ 10 $    & $ 2.241922364228 \, (2.242) $ \quad\quad & $ 2.683607543289  \, (2.684) $ \\
  $ 100 $   & $ 2.926472047450 \, (2.926) $ \quad\quad & $ 3.061894259945  \, (3.061) $ \\
  $ 1000 $  & $ 3.118751713938 \, (3.118) $ \quad\quad & $ 3.150204534073  \, (3.150) $\\
  \hline\hline
 \multicolumn{3}{c}{$ x_{\rm e}(N_p = 4.54) = 3.174901573278$}\\
  \hline\hline
\end{tabular}
\end{center}
\label{tab-roots_1}
\end{table} 

\begin{figure*}[htpb!]
   \includegraphics[width=0.5\textwidth]{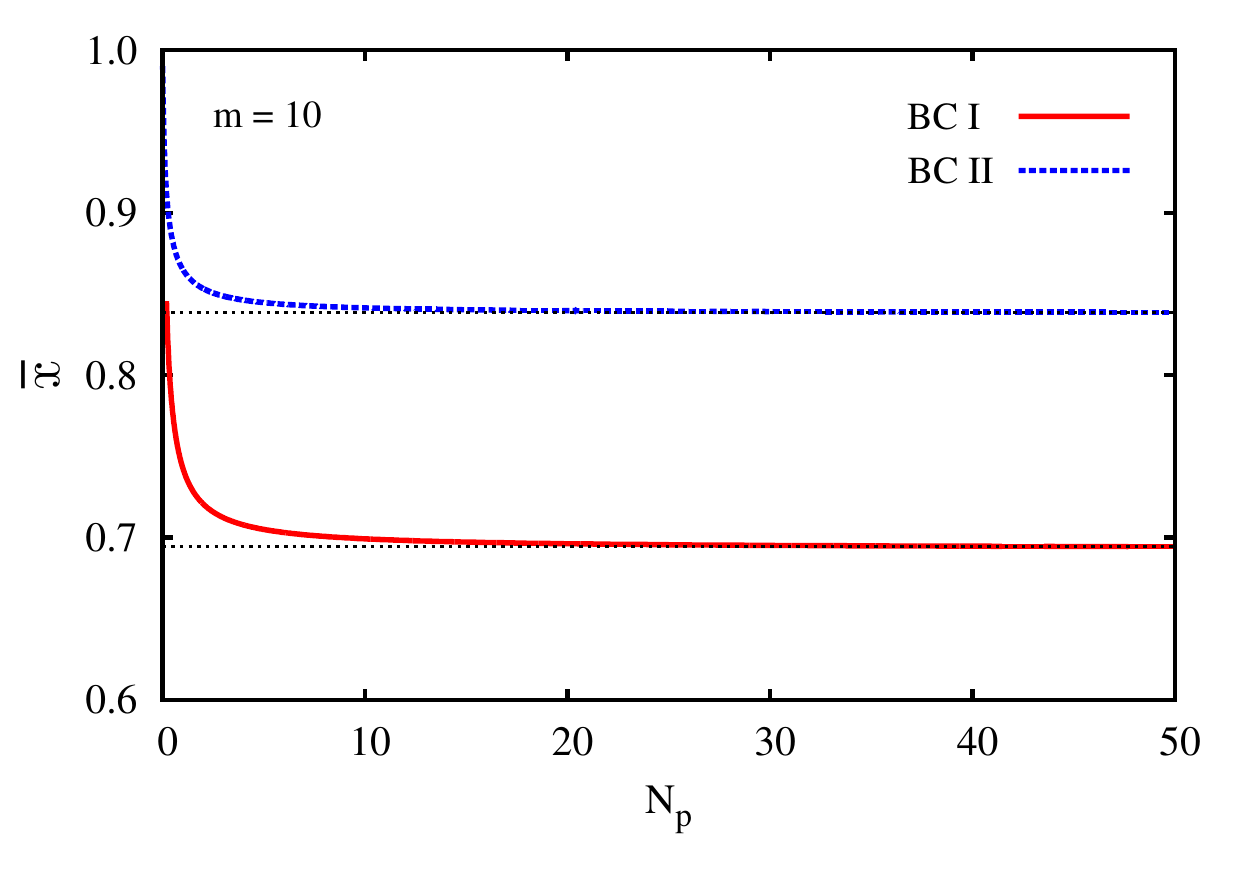}\includegraphics[width=0.5\textwidth]{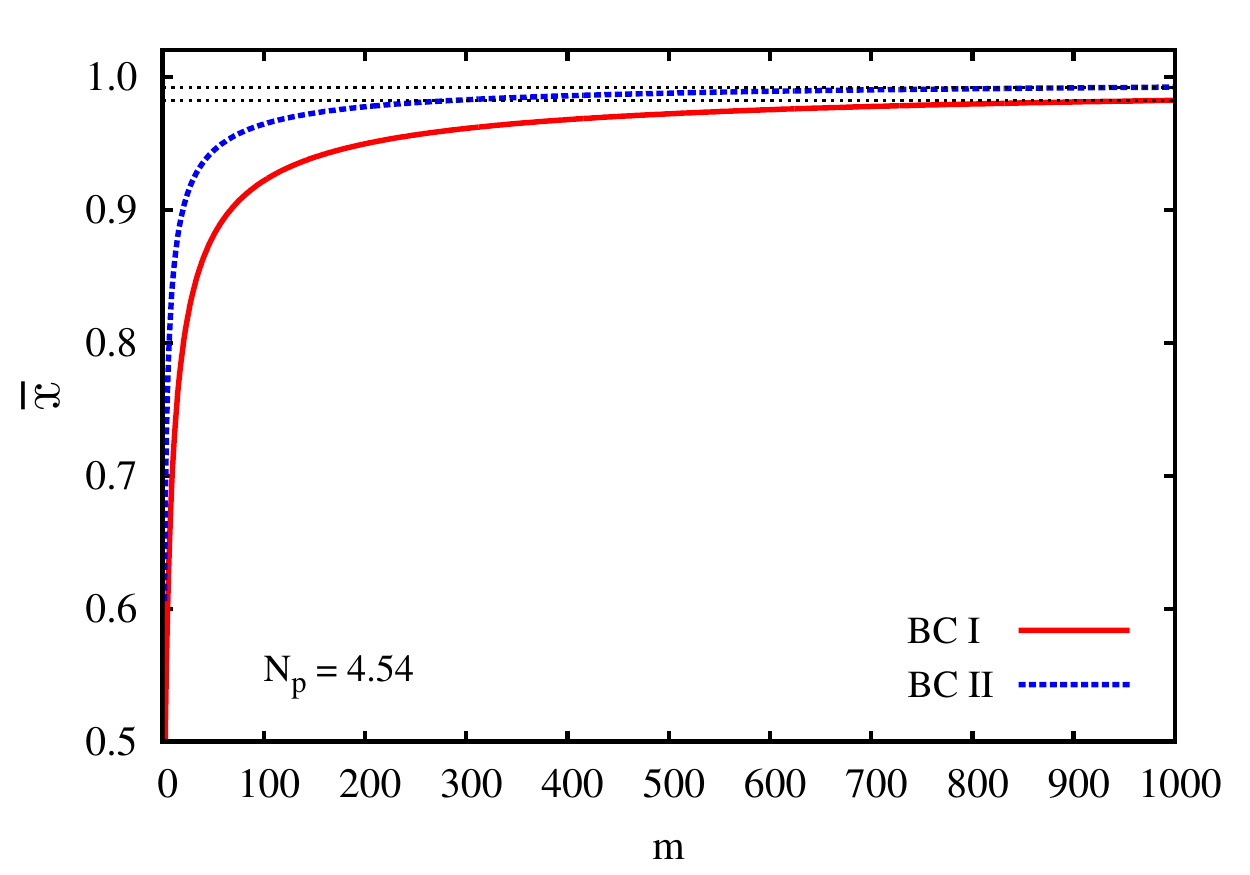}
  \caption{{\it Left plots}: Estimates of $\bar{x}$, as a function of polytropic index $N_{\rm p}$ and for azimuthal number $m=10$, for the boundary conditions {\rm BC\,I} and {\rm BC\,II}. {\it Right plots}: Estimates of $\bar{x}$, as a function of azimuthal number $m$ and polytropic index $N_{\rm p}=4.54$, for the boundary conditions {\rm BC\,I} and {\rm BC\,II}. The results were obtained numerically using a root-finding algorithm.}
   \label{x_min_BCI_II}
  \end{figure*}

In Fig.~\ref{x_min_BCI_II} we plot the estimates of $\bar{x} \equiv r_{\rm min}^{(0)}/r_{\rm e}$, as a function of polytropic index $N_{\rm p}$ (for the azimuthal number $m=10$) and as a function of azimuthal number $m$ (for the polytropic index $N_{\rm p}=4.54$). Essentially, the growth of the polytropic index $N_{\rm p}$ implies in a decrease of $\bar{x}$. On the other hand, the growth of the azimuthal number $m$ implies in a growth of $\bar{x}$. For large values of the polytropic index $N_{\rm p}$, $\bar{x}$ becomes constant ($\bar{x}<1$) and for large values of the azimuthal number $m$, $\bar{x}\rightarrow1$, denoting that the static resonances ($\omega=0$) occur at $r_{\rm min}=r_{\rm e}$, for large values of the azimuthal number $m$.

\section{Conclusion} 
\label{sec-Conclusion}
Here we used the method of lines (MOL), direct integration (DI) and continued-fraction (CF) methods to obtain the QNM frequencies of the polytropic hydrodynamic vortex, an effective spacetime produced by purely circulating compressible fluid that satisfies a polytropic equation of state. To validate our results, we compared the QNM frequencies obtained via these three different methods and obtained excellent agreement. Furthermore, we studied the polytropic hydrodynamic vortex in the regime between stability and instability, obtaining the configuration of the parameters of this system (azimuthal number $m$ and polytropic index $N_P$) required for the onset of the ergoregion instability. We focused our attention on the dependence of the QNM frequencies with the polytropic index $N_{\rm p}$. As the polytropic index $N_{\rm p}$ increases, the magnitude of real and imaginary parts of the QNM frequencies increase (decrease) for unstable (stable) modes. Furthermore, we have shown that the polytropic hydrodynamic vortex, a compressible system with an ergoregion and without an event horizon is unstable. Together with the instability of the corresponding incompressible system, namely, the hydrodynamic vortex studied in Ref.~\cite{Oliveira:2014oja}, this establishes the ergoregion instability as a generic phenomena. The instability is more clearly revealed by computing QNMs, imposing boundary conditions at $r=r_{\rm min}$ located inside the outer boundary $r_{\rm e}$ of the ergoregion. Finally, we have shown that the onset of ergoregion instability approaches the outer boundary of the ergoregion as the azimuthal number $m$ increases, denoting that for large values of the azimuthal number $m$ the QNMs become unstable, independently of the configuration (circulation $C$ and polytropic index $N_{\rm p}$) of the system (as may be seen in Fig.~\ref{Freq_rmin_BCI_II}\,). Furthermore, we have shown that for large values of azimuthal number $m$, the static resonances ($\omega=0$) occur at $r_{\rm min}=r_{\rm e}$ (as may be seen in Table~\ref{tab-roots_1} and Fig.~\ref{x_min_BCI_II}).

\begin{acknowledgments}
V. C. thanks the Universidade Federal do Par\'a (UFPA) in
Bel\'em for the kind hospitality.
The authors would like to thank Conselho Nacional de Desenvolvimento 
Cient\'\i fico e Tecnol\'ogico (CNPq), Coordena\c{c}\~ao 
de Aperfei\c{c}oamento de Pessoal
de N\'\i vel Superior (CAPES) and Funda\c{c}\~ao Amaz\^onia Paraense de Amparo \`a Pesquisa (FAPESPA) for partial financial support.
We acknowledge financial support provided under the European
Union's FP7 ERC Starting Grant ``The dynamics of black holes: testing
the limits of Einstein's theory'' grant agreement no. DyBHo--256667 and the
H2020 ERC Consolidator Grant ``Matter and strong-field gravity: New frontiers in Einstein's theory'' grant agreement no. MaGRaTh--646597.
This research was supported in part by Perimeter Institute for Theoretical Physics. 
Research at Perimeter Institute is supported by the Government of Canada through 
Industry Canada and by the Province of Ontario through the Ministry of Economic Development 
$\&$ Innovation.
This work was also supported by the NRHEP 295189 FP7-PEOPLE-2011-IRSES Grant, and by FCT-Portugal through projects
CERN/FP/123593/2011. 

\end{acknowledgments}



\begin{thebibliography}{99}
  
\bibitem{Unruh:1980cg} 
  W.~G.~Unruh,
  Experimental black hole evaporation,
  Phys.\ Rev.\ Lett.\  {\bf 46}, 1351 (1981).
  
\bibitem{Visser:1997ux} 
  M.~Visser,
  Acoustic black holes: Horizons, ergospheres, and Hawking radiation,
  Class.\ Quant.\ Grav.\  {\bf 15}, 1767 (1998)
  [gr-qc/9712010].
  
\bibitem{AMproc} V. Cardoso, L. C. B. Crispino, S. Liberati, E. S. Oliveira and M. Visser (Eds.),
{\it Analogue spacetimes: the first thirty years} (Editora Livraria da F\'{\i}sica, S\~ao Paulo, 2013).    
    
\bibitem{Crispino:2007zz}
  L.~C.~B.~Crispino, E.~S.~Oliveira and G.~E.~A.~Matsas,
  Absorption cross section of canonical acoustic holes,
  Phys.\ Rev.\ D {\bf 76} (2007) 107502.

\bibitem{Oliveira:2010zzb}
  E.~S.~Oliveira, S.~R.~Dolan and L.~C.~B.~Crispino,
  Absorption of planar waves in a draining bathtub,
  Phys.\ Rev.\ D {\bf 81} (2010) 124013.

\bibitem{Dolan:2009zza}
  S.~R.~Dolan, E.~S.~Oliveira and L.~C.~B.~Crispino,
  Scattering of sound waves by a canonical acoustic hole,
  Phys.\ Rev.\ D {\bf 79} (2009) 064014
  [arXiv:0904.0010 [gr-qc]].

  
\bibitem{Dolan:2011zza}
  S.~R.~Dolan, E.~S.~Oliveira and L.~C.~B.~Crispino,
  Aharonov-Bohm effect in a draining bathtub vortex,
  Phys.\ Lett.\ B {\bf 701} (2011) 485.
  
\bibitem{Dolan:2012yc} 
  S.~R.~Dolan and E.~S.~Oliveira,
  Scattering by a draining bathtub vortex,
  Phys.\ Rev.\ D {\bf 87}, 124038 (2013)
  [arXiv:1211.3751 [gr-qc]].

\bibitem{Barcelo:2005fc} 
  C.~Barcelo, S.~Liberati and M.~Visser,
  Analogue gravity,
  Living Rev.\ Rel.\  {\bf 8}, 12 (2005)
  [Living Rev.\ Rel.\  {\bf 14}, 3 (2011)]
  [gr-qc/0505065].

\bibitem{Brito:2015oca} 
  R.~Brito, V.~Cardoso and P.~Pani,
  Superradiance,
  arXiv:1501.06570 [gr-qc].
  
\bibitem{Visser:2004zs} 
  M.~Visser and S.~E.~C.~Weinfurtner,
  Vortex geometry for the equatorial slice of the Kerr black hole,
  Class.\ Quant.\ Grav.\  {\bf 22}, 2493 (2005)
  [gr-qc/0409014].
  
\bibitem{Slatyer:2005ty} 
  T.~R.~Slatyer and C.~M.~Savage,
  Superradiant scattering from a hydrodynamic vortex,
  Class.\ Quant.\ Grav.\  {\bf 22}, 3833 (2005)
  [cond-mat/0501182].
 
\bibitem{Fischer:2001jz} 
  U.~R.~Fischer and M.~Visser,
  Riemannian geometry of irrotational vortex acoustics,
  Phys.\ Rev.\ Lett.\  {\bf 88}, 110201 (2002)
  [cond-mat/0110211].

\bibitem{no-horizon}
Here, the choice of a purely circulating system and not a system that includes a radial flow as, e.g., the draining bathtub, has been made: (i) to isolate the effect (ergoregion instability), that is only related to the ergoregion and (ii) to show that there is a simple experimental setup where the instability exists.

\bibitem{Friedman:1978wla} 
  J.~L.~Friedman,
  Generic instability of rotating relativistic stars,
  Commun.\ Math.\ Phys.\  {\bf 62}, no. 3, 247 (1978).
  
\bibitem{Oliveira:2014oja} 
  L.~A.~Oliveira, V.~Cardoso and L.~C.~B.~Crispino,
  Ergoregion instability: The hydrodynamic vortex,
  Phys.\ Rev.\ D {\bf 89}, 124008 (2014)
  [arXiv:1405.4038 [gr-qc]].
  
\bibitem{Cherubini:2013iea}
  C.~Cherubini and S.~Filippi,
  Classical field theory of the Von Mises equation for irrotational polytropic inviscid fluids,
  J.\ Phys.\ A {\bf 46} (2013) 115501.
  
\bibitem{Cherubini:2011zza} 
  C.~Cherubini and S.~Filippi,
  Acoustic metric of the compressible draining bathtub,
  Phys.\ Rev.\ D {\bf 84}, 084027 (2011).

 \bibitem{Horedt}   
  G.~P.~Horedt, Polytropes, {\it Applications in Astrophysics and
Related Fields} (Kluwer Academic Publishers, Dordrecht,
The Netherlands, 2004).

\bibitem{Turns}
S.~R.~Turns, {\it Thermal-Fluid Sciences: An Integrated Approach} (Cambridge University Press, Cambridge, England, 2006).

\bibitem{non-physical} 
The term ``non-physical'' here is use to denote to fact that the Kretschmann invariant goes to infinite at $r\rightarrow r_{\rm c}$, i.e., the polytropic hydrodynamic vortex has an essential singularity at critical radius $r=r_{\rm c}$ for an polytropic index $N_{\rm p}>-3/2$ (with $N_{\rm p} \neq -1$).

\bibitem{Turrell}
G.~Turrell, {\it Gas Dynamics: Theory and Applications} (John Willey \& Sons, Chichester, England, 1997). 

\bibitem{Laplace}
In 1816, Pierre-Simon Laplace explained that the propagation of sound waves in a gas is an adiabatic process due to the fact the speed of vibration of sound waves is a process rapid enough such that occurs without transfer of heat throughout the process~\cite{Turrell}. 

\bibitem{Perry}
R.~H.~Perry and D.~W.~Green, {\it Perry's Chemical Engineers' Handbook} (McGraw-Hill Professional, New York, USA, 1997).

\bibitem{Lax:1948} 
  M.~Lax and H.~Feshbach,
  Absorption and Scattering for Impedance Boundary Conditions on Spheres and Circular Cylinders,
  J.\ Acoust.\ Soc.\ Am. {\bf 20}, 108 (1948).
  
\bibitem{Rinne}
O.~Rinne, Ph.D. thesis, University of Cambridge (2006) [arXiv:gr-qc/0601064].
  
\bibitem{Witek:2012tr} 
  H.~Witek, V.~Cardoso, A.~Ishibashi, and U.~Sperhake,
  Superradiant instabilities in astrophysical systems,
  Phys.\ Rev.\ D {\bf 87}, 043513 (2013)
  [arXiv:1212.0551 [gr-qc]].
  
  
\bibitem{Dolan:2011ti} 
  S.~R.~Dolan, L.~A.~Oliveira, and L.~C.~B.~Crispino,
  Resonances of a rotating black hole analogue,
  Phys.\ Rev.\ D {\bf 85}, 044031 (2012)
  [arXiv:1105.1795 [gr-qc]].
  
\bibitem{Dolan:2012yt} 
  S.~R.~Dolan,
  Superradiant instabilities of rotating black holes in the time domain,
  Phys.\ Rev.\ D {\bf 87}, 124026 (2013)
  [arXiv:1212.1477 [gr-qc]].
  
  
\bibitem{Butcher}
J.~C.~Butcher, {\it Numerical Methods for Ordinary Differential Equations} (John Willey \& Sons, Chichester, England, 2003).  

\bibitem{Dolan:2010zza} 
  S.~R.~Dolan, L.~A.~Oliveira and L.~C.~B.~Crispino,
  Quasinormal modes and Regge poles of the canonical acoustic hole,
  Phys.\ Rev.\ D {\bf 82}, 084037 (2010)
  [arXiv:1407.3904 [gr-qc]].
  
\bibitem{Leaver:1985ax} 
  E.~W.~Leaver,
  An Analytic representation for the quasi normal modes of Kerr black holes,
  Proc.\ Roy.\ Soc.\ Lond.\ A {\bf 402}, 285 (1985).
  
\bibitem{Onozawa:1995vu} 
  H.~Onozawa, T.~Mishima, T.~Okamura, and H.~Ishihara,
  Quasinormal modes of maximally charged black holes,
  Phys.\ Rev.\ D {\bf 53}, 7033 (1996).
  
\bibitem{Berti:2009kk} 
  E.~Berti, V.~Cardoso and A.~O.~Starinets,
  Quasinormal modes of black holes and black branes,
  Class.\ Quant.\ Grav.\  {\bf 26}, 163001 (2009)
  [arXiv:0905.2975 [gr-qc]].
  

\bibitem{Hod:2014hda} 
  S.~Hod,
  Onset of superradiant instabilities in the hydrodynamic vortex model,
  Phys.\ Rev.\ D {\bf 90}, 027501 (2014)
  [arXiv:1405.7702 [gr-qc]].
  
\bibitem{Marecki} 
  P.~Marecki and R.~ Sch{\"u}tzhold,
  Whispering gallery like modes along pinned vortices,
  JETP Letters {\bf 96}, 674 (2012).
  
  
  
  \bibitem{Grad}
I. S. Gradshteyn and I. M. Ryzhik, {\it Tables of Integrals,
Series, and Products} (Academic Press, New York, 1980).


 
\end{thebibliography}
\end{document}